\documentclass[10pt]{article}

\usepackage{amsmath}

\usepackage{array}

\usepackage{appendix}

\usepackage{tocloft}                   

\usepackage{graphicx}

\usepackage{amsfonts}

\usepackage{amssymb}

\usepackage{mathrsfs}

\usepackage{yfonts}

\usepackage{euscript}

\usepackage{centernot}                 

\usepackage{ifsym}                     

\usepackage{lmodern}                   

\usepackage{upgreek}

\usepackage{mathtools}

\usepackage{bm}

\usepackage{undertilde}

\usepackage{dutchcal}

\usepackage{amsfonts}

\usepackage{amssymb}

\usepackage{mathrsfs}

\usepackage{upgreek}

\usepackage{mathtools}

\usepackage{color}

\usepackage{slantsc}
\usepackage{calligra}

\usepackage{bbold}          

\usepackage[T1]{fontenc}

\usepackage{epsf}

\usepackage{latexsym}

\usepackage{tipa}

\usepackage{makeidx}

\makeindex

\def\n{\noindent}

\def\m{\mbox{ }}

\def\be{\begin{equation}}                

\def\ee{\end{equation}}

\def\beq{\begin{equation}}                   

\def\eeq{\end{equation}}

\def\bea{\begin{eqnarray}}                   

\def\eea{\end{eqnarray}}

\def\mma {\m , \m \m }                       

\def\mmc {\m : \m \m }                       

\def\mmp {\m + \m}                       

\def\mmm {\m - \m}                       

\def\mapprox {\m \approx \m}                       

\def\msim {\m \approx \m}                       

\def\mlongrightarrow {\m \longrightarrow \m}

\def\msubseteq {\m \subseteq \m}

\def\msupseteq {\m \supseteq \m}

\def\!{\hspace{-1.6667em}}                   

\def\s{\stackrel}                            

\def\Proof{{\n{\u{Proof}}}\m}

\def\u{\underline}       

\def\uc{\underbracket}   
\def\uo{\utilde}         

\def\half{\mbox{$\frac{1}{2}$}}

\def\lp{l^{\prime}}

\def\bia{\mbox{\boldmath$a$}}

\def\sbig{\mbox{\scriptsize\boldmath$g$}}

\def\bin{\mbox{\boldmath$n$}}

\def\bip{\mbox{\boldmath$p$}}

\def\biu{\mbox{\boldmath$u$}}

\def\biA{\mbox{\boldmath$A$}}

\def\biB{\mbox{\boldmath$B$}}

\def\biC{\mbox{\boldmath$C$}}              

\def\biD{\mbox{\boldmath$D$}}

\def\biG{\mbox{\boldmath  $G$}}             %
\def\biF{\mbox{\boldmath $F$}}             %

\def\biH{\mbox{\boldmath$H$}}

\def\biP{\mbox{\boldmath$P$}}

\def\biQ{\mbox{\boldmath$Q$}}

\def\biS{\mbox{\boldmath$S$}}

\def\biW{\mbox{\boldmath$W$}}

\def\bttf{\mbox{\ttfamily\fontseries{b}\selectfont f}}                   %

\def\bttA{\mbox{\ttfamily\fontseries{b}\selectfont A}}                   %

\def\btta{\mbox{\ttfamily\fontseries{b}\selectfont a}}                   %

\def\sbiU{\mbox{\ttfamily\fontseries{b}\selectfont U}}                   %

\def\sbiC{\mbox{\ttfamily\fontseries{b}\selectfont C}}

\def\bTheta{\mbox{\boldmath$\Theta$}}

\def\bphi{\mbox{\boldmath$\phi$}} 

\def\balpha{\mbox{\boldmath$\alpha$}} 

\def\sbalpha{\mbox{\scriptsize\boldmath$\alpha$}} 

\def\sbupxi{\mbox{\scriptsize\boldmath$\upxi$}}

\def\mC{\mbox{C}}                        

\def\mH{\mbox{H}} 

\def\mN{\mbox{N}}

\def\mg{\mbox{g}}

\def\mo{\mbox{o}}

\def\mp{\mbox{p}}

\def\bh{\u{\u{\mbox{h}}}  }            

\def\bX{\mbox{\bf X}}

\def\bY{\mbox{\bf Y}}

\def\bZ{\mbox{\bf Z}}

\def\bee{\mbox{\bf e}}             

\def\bh{\mbox{\bf h}}

\def\btheta{\mbox{\boldmath$\theta$}}             

\def\bupSigma{\mbox{\boldmath$\Sigma$}}                 
\def\bupxi{\mbox{\boldmath$\xi$}}                       

\def\bupchi{\mbox{\boldmath$\chi$}}                     

\def\bcalS{\mbox{\boldmath ${\cal S}$}}

\def\bcalD{\mbox{\boldmath ${\cal D}$}}

\def\fg{\mbox{\tt g}}                          

\def\fo{\mbox{\sffamily o}}

\def\fz{\mbox{\sffamily z}}

\def\fC{\mbox{\sffamily C}}

\def\fE{\mbox{\sffamily E}}

\def\fF{\mbox{\sffamily F}}

\def\fI{\mbox{\sffamily I}}

\def\fO{\mbox{\sffamily O}}

\def\fP{\mbox{\sffamily P}}

\def\fS{\mbox{\sffamily S}}

\def\fZ{\mbox{\sffamily Z}}

\def\sb{\mbox{\scriptsize b}}

\def\se{\mbox{\scriptsize e}}

\def\sf{\mbox{\scriptsize f}}

\def\sg{\mbox{\scriptsize g}}

\def\si{\mbox{\scriptsize i}}

\def\sn{\mbox{\scriptsize n}} 

\def\so{\mbox{\scriptsize o}}

\def\sss{\mbox{\scriptsize s}}  

\def\sE{\mbox{\scriptsize E}}

\def\sG{\mbox{\scriptsize G}}

\def\sN{\mbox{\scriptsize N}}

\def\sS{\mbox{\scriptsize S}}

\def\sT{\mbox{\scriptsize T}}

\def\sV{\mbox{\scriptsize V}}

\def\sfI{\mbox{\sffamily{\scriptsize I}}}      

\def\sfP{\mbox{\sffamily{\scriptsize P}}}      

\def\sbttO{\mbox{\scriptsize\boldmath{\tt O}}}

\def\sbcC{\mbox{\boldmath \scriptsize ${\cal C}$}}

\def\sbcF{\mbox{\boldmath \scriptsize ${\cal F}$}}

\def\sbcG{\mbox{\boldmath \scriptsize ${\cal G}$}}

\def\sbcN{\mbox{\boldmath \scriptsize ${\cal N}$}}

\def\sbcS{\mbox{\boldmath \scriptsize ${\cal S}$}}

\def\bscc{\mbox{{\boldmath \scriptsize${\cal c}$}}}                               

\def\bsct{\mbox{{\boldmath \scriptsize${\cal t}$}}}                               

\def\bscJ{\mbox{{\boldmath \scriptsize${\cal J}$}}}                               

\def\bscT{\mbox{{\boldmath \scriptsize${\cal T}$}}}                               

\def\bscg{\mbox{{\boldmath ${\cal g}$}}}                               

\def\sbcg{\mbox{{\boldmath ${\cal g}$}}}                               

\def\bsch{\mbox{{\boldmath ${\cal h}$}}}                               

\def\bscD{\mbox{{\boldmath \scriptsize${\cal D}$}}}                               

\def\bscZ{\mbox{{\boldmath \scriptsize${\cal Z}$}}}                               

\def\bscF{\mbox{{\boldmath \scriptsize${\cal F}$}}}                               

\def\bscP{\mbox{\boldmath\scriptsize${\cal P}$}}                               

\def\bscS{\mbox{\boldmath \scriptsize${\cal S}$}}                               

\def\bscI{\mbox{{\boldmath \scriptsize${\cal I}$}}}                               

\def\bscM{\mbox{{\boldmath \scriptsize${\cal M}$}}}  

\def\bscz{\mbox{{\boldmath \scriptsize${\cal z}$}}}

\def\bttJ{\mbox{\boldmath {\tt J}}}

\def\btto{\mbox{\boldmath {\tt o}}}

\def\Ob{\mbox{\boldmath {\tt O}}}         

\def\BOb{\mbox{\boldmath {\tt B}}}

\def\ob{\mbox{\boldmath {\tt o}}}         

\def\uob{\mbox{\boldmath {\tt u}}}

\def\sumi2{\sum\mbox{}_{\mbox{}_{\mbox{\scriptsize $i$=1}}}^2}

\def\sumi3{\sum\mbox{}_{\mbox{}_{\mbox{\scriptsize $i$=1}}}^3}

\def\sumkn{\sum\mbox{}_{\mbox{}_{\mbox{\scriptsize $k$=1}}}^{n}}

\def\sumABcycles3{\sum\mbox{}_{\mbox{}_{\mbox{\scriptsize cycles $A,B$=1}}}^{3}}

\def\sumCDcycles3{\sum\mbox{}_{\mbox{}_{\mbox{\scriptsize cycles $C,D$=1}}}^{3}}

\def\sumj3{\sum\mbox{}_{\mbox{}_{\mbox{\scriptsize $j$=1}}}^3}

\def\sumk3{\sum\mbox{}_{\mbox{}_{\mbox{\scriptsize $k$=1}}}^3}

\def\prodiA1{\prod\mbox{}_{\mbox{}_{\mbox{\scriptsize $i$=1}}}^{A - 1}}

\def\bigtimes{\mbox{\Large $\times$}}

\def\d{\textrm{d}}                                                  

\def\pa{\partial}                                                   

\def\bpa{\mbox{\boldmath$\partial$}}                                                   

\def\Lie{\negthickspace\negthickspace\negthickspace{\cal L}}        

\def\es{\m = \m}

\def\:={\m := \m}

\def\=:{\m =: \m}

\def\ls{\m < \m}

\def\FrT{\mathfrak{T}}                                         

\def\FrC{\mbox{$\mathfrak{C}$}}

\def\FrX{\mathfrak{X}}                                         
\def\lFrs{\mbox{$\mathfrak{S}$}}                         
\def\LFrS{\mbox{$\mathfrak{S}$}}                               %

\def\FrU{\mbox{$\mathfrak{U}$}}                                
\def\FrV{\mbox{$\mathfrak{V}$}}                                
\def\FrS{\mbox{$\mathfrak{S}$}}    

\def\Frm{\mbox{\Large $\mathfrak{m}$}}                         
\def\FrM{\mbox{$\mathfrak{M}$}}                                
\def\lFrs{\mathfrak{S}}                                        

\def\lFrg{\mbox{\Large$\mathfrak{g}$}}                         
\def\sFrg{\mbox{\footnotesize$\mathfrak{g}$}}                  
\def\Frg{\mbox{\normalsize $\mathfrak{g}$}}                    

\def\FrB{\mbox{$\mathfrak{B}$}}                                

\def\FrF{\mbox{\boldmath$\mathfrak{F}$}}                       

\def\FrT{\mbox{\boldmath$\mathfrak{T}$}}                       
\def\bFrPP{\mbox{\boldmath$\mathfrak{P}$}}                     
 
\def\sFrf{\mbox{\large $\mathfrak{f}$}}                          

\def\FrG{\mathfrak{G}}                                         
\def\NH{\u{\,H\,}}                                             

\def\Hilb{\mbox{{\boldmath$\mathfrak{H}$}ilb}}                 
\def\Frc{\mbox{\Large $\mathfrak{c}$}}                         
\def\cb{\mbox{\bf ,} \,}                                                     

\def\ll{\mbox{\bf [} \,}                                                    

\def\ls{\mbox{\bf |[} \,}                                                     

\def\lp{\mbox{\bf \{} \,}                                                    

\def\rl{\, \mbox{\bf ]}}                                                    

\def\rs{\, \mbox{\bf ]|}}                                                     

\def\rp{\, \mbox{\bf \}}}                                                    

\def\rd{\, \mbox{\bf \}} \mbox{}^{\mbox{\bf *}}}                             

\def\rV{\, \mbox{\bf ]|}_{\sV}}                                              

\def\rG{\, \mbox{\bf ]|}_{\sG}}                                              

\def\gen{\bscg\mbox{en}}                                                    

\def\top{\bsct\mbox{op}}                                               

\def\flin{\bscf\mbox{lin}}                                               

\def\gauge{\bscg\mbox{auge}}                                             

\def\sgen{\bscg\se\sn}                                                      

\def\FlinGen{\sFrf\mbox{lin-}\Frg\mbox{en}}                                  

\def\GaugeGen{\Frg\mbox{auge-}\Frg\mbox{en}}                                

\def\sGen{\sFrg\se\sn}                                                      

\def\con{\mbox{{\boldmath \scriptsize${\cal C}$}}}                           

\def\scon{\bscc\so\sn}                                                       

\def\sCon{\sFrC\so\sn}                                                     

\def\TopCon{\FrT\mbox{op-}\FrC\mbox{on}}                                  

\def\FlinCon{\FrF\mbox{lin-}\FrC\mbox{on}}                                  

\def\GaugeCon{\FrG\mbox{auge-}\FrC\mbox{on}}                                

\def\sCon{\sFrC\so\sn}                                                     

\def\LattCon{\lattice_{\sCon}}                                                

\def\gaugeobs{\btto}

\def\sobs{\sbttO}

\def\Obs{\FrO\mbox{bs}}                                                     

\def\BiObs{\FrB\mbox{iObs}}                                                 

\def\UnresObs{\FrU\mbox{nres-}\FrO\mbox{bs}}                                

\def\TopObs{\FrT\mbox{op-}\FrO\mbox{bs}}                                  %

\def\GaugeObs{\FrG\mbox{auge-}\FrO\mbox{bs}}                                %

\def\FlinObs{\FrF\mbox{lin-}\FrO\mbox{bs}}                                  %

\def\LattObs{\lattice_{\tFrO\sb\sss}}                                                

\def\scC{\mbox{\scriptsize ${\cal C}$}}                    
\def\bscf{\mbox{\boldmath ${\cal f}$}}

\def\scH{\mbox{\scriptsize ${\cal H}$}}                    

\def\bscP{\mbox{\boldmath\scriptsize ${\cal P}$}}

\def\bscP{\mbox{\boldmath\scriptsize ${\cal P}$}}

\def\bscS{\mbox{\boldmath\scriptsize ${\cal S}$}}

\def\scS{\mbox{\scriptsize ${\cal S}$}}                    

\def\Sec{\bscS\mbox{\bf e}}                                

\def\flin{\bscf\mbox{lin}}                                  

\def\bFlin{\sbcF\mbox{\bf lin}} 

\def\Chronos{\scC\mbox{hronos}}                            

\def\bGauge{\sbcG\mbox{\bf auge}} 

\def\NSC{\sbcN\sbcS\sbcC}                                     

\def\NSFlin{\sbcN\sbcS\sbcF\mbox{\bf lin}}                        

\def\chronos{\mbox{\ttfamily\fontseries{b}\selectfont C}} 
					   
\def\Dirac{\mbox{\ttfamily\fontseries{b}\selectfont D}}    
														   
\def\gauge{\mbox{\ttfamily\fontseries{b}\selectfont G}}                  

\def\Kuchar{\mbox{\ttfamily\fontseries{b}\selectfont K}}                  
\def\bNsD{\mbox{\ttfamily\fontseries{b}\selectfont NSD}}                          

\def\bNsK{\mbox{\ttfamily\fontseries{b}\selectfont NSK}}                          

\def\NSFlin{\sbcN\sbcS\sbcF\mbox{\bf lin}}                 

\def\Unres{\mbox{\ttfamily\fontseries{b}\selectfont U}}                   %

\def\Top{\mbox{\ttfamily\fontseries{b}\selectfont T}}                   %

\def\unres{\mbox{\ttfamily\fontseries{b}\selectfont u}}
                                 
\def\bFrC{\mbox{\boldmath$\mathfrak{C}$}}                            

\def\bFrg{\mbox{\boldmath$\mathfrak{g}$}}                            
 
\def\Phase{\mbox{{\boldmath$\mathfrak{P}$}hase}}                     

\def\bFrR{\mbox{\boldmath$\mathfrak{R}$}}                            
\def\Rig-Phase{\bFrR\mbox{ig-}\Phase}                                
                                                                													   
\def\bFro{\mbox{\Large $\mathfrak{o}$}}                              
\def\tFrO{\mbox{\tiny\boldmath$\mathfrak{O}$}} 
 
\def\lFra{\mbox{\Large $\mathfrak{a}$}} 		                     

\def\bFrZ{\mbox{\boldmath$\mathfrak{Z}$}}                            

\def\bFrR{\mbox{\boldmath$\mathfrak{R}$}}                            

\def\bFrR{\mbox{\boldmath$\mathfrak{R}$}}                            

\def\1mat{\u{\u{1}}}                                                 

\def\Positive-Modespace{\mbox{{\boldmath$\mathfrak{M}$}odespace$^+$}}

\def\POSITIVE-MODESPACE{\mbox{{\boldmath$\mathfrak{M}$}ODESPACE$^+$}}

\def\PRiem{\bFrPP\bFrR\mbox{iem}}                                    

\def\FrO{\mbox{$\mathfrak{O}$}}                                      

\def\bFrO{\mbox{\boldmath$\mathfrak{O}$}}                            

\def\Top{\FrT\mo\mp}

\def\lattice{\mbox{\bf\Large$\mathfrak{L}$}}                                      

\def\sFrC{\mbox{\boldmath\scriptsize$\mathfrak{C}$}}                        

\def\Kin-Hilb{\mbox{{\boldmath$\mathfrak{K}$}in-\Hilb}}                     

\def\Mid-Hilb{\mbox{{\boldmath$\mathfrak{M}$}id-\Hilb}}                     

\def\Dyn-Hilb{\mbox{{\boldmath$\mathfrak{D}$}yn-\Hilb}}                     

\def\5Star{\mbox{\Large$\star$}}              

\def\K{Kucha\v{r} }

\def\peq{\m \mbox{`='} \m}

\def\peqs{\m \m \mbox{`='} \m \m}

\def\npeq{\m \mbox{`$\neq$'} \m}

\begin{document}

\begin{center}

\Large{\bf Nambu variant of Local Resolution of Problem of Time} 

\Large{\bf and Background Independence}

\vspace{.1in}

{\large \bf Edward Anderson}$^1$ 

\end{center}

\begin{abstract}

A Local Resolution of the Problem of Time has recently been given, alongside reformulation as A Local Theory of Background Independence.   
The classical part of this can be viewed as requiring just Lie's Mathematics, 
albeit entrenched in subsequent topological and differential-geometric developments and extended to contemporary Physics' state spaces.
We now widen this approach by mild recategorization to one based on Nambu's generalization of Lie's Mathematics, as follows. 
i)  In this approach, the Lie derivative still suffices to encode Relationalism. 
ii) Closure is now assessed using the Nambu bracket -- with $n$ slots rather than 2, so the first nontrivially Lie case has 3 slots -- 
and a `Nambu Algorithm' analogue of the Dirac and Lie Algorithms.  
This produces a class of Nambu algebraic structures of generators or of first-class constraints.   
iii) Nambu observables are defined by Nambu brackets zero-commutation with generators or with first-class constraints;
we use the Nambu analogue of the Jacobi identity to simplify this discussion relative to a previous treatment. 
These Nambu brackets relations can moreover be recast as explicit PDEs to be solved using the Flow Method. 
Nambu observables themselves form Nambu algebras. 
Lattices of Nambu constraint or generator algebraic substructures 
furthermore induce dual lattices of Nambu observables subalgebras.  
iv) Deformation of Nambu algebraic structures encountering Rigidity gives a means of Constructing more structure from less.  
v) Reallocation of Intermediary-Object Invariance gives the general Nambu algebraic structure's analogue of posing Refoliation Invariance for GR. 
We also draw some motivation from M-Theory's use of Nambu Mathematics along the lines of Bagger, Lambert and Gustavsson,   
finding some qualitative distinctions between this, GR and Supergravity as regards how Background Independence is realized.

\end{abstract}

$^1$ dr.e.anderson.maths.physics *at* protonmail.com

\section{Introduction}\label{Introduction}

It has been recently demonstrated that \cite{ABook, ALett, DO-1, I, II, III, IV, V, VI, VII, VIII, Higher-Lie, IX, X, XI, XII, XIII, XIV} 
Lie's Mathematics \cite{Lie, Yano55, Jacobson, Serre-Lie, NR66, Yano70, Gilmore, Olver2, CM, M08, Lee2, Olver, BCHall} 
suffices to construct A Local Resolution of the Problem of Time (ALRoPoT) \cite{Battelle, DeWitt67, Dirac, K92, I93, APoT, APoT2, ABook}, 
which in turn can be reformulated as \cite{ABook, A-CBI} A Local Theory of Background Independence (ALToBI). 

\m 

\n This locally-smooth approach is moreover sufficiently well-defined to extend to various other (at least locally) differential-geometric structures. 
The purpose of the current Article is to outline one of the more interesting cases: the `Nambu Mathematics' 
\cite{Nambu, BF75, Filippov, T94, V98, GM98, CZ02, odd-even, Nambu-Rev} counterpart. 
See \cite{Nijenhuis} for use instead of `Nijenhuis Mathematics' \cite{S40-53-N55, FN56, NR66}, 
while \cite{ABook, XIV} already made mention of the simpler graded, alias supersymmetric, extension.    

\m
 
\n Sec \ref{NM} presents Nambu Mathematics' primary distinctive structures. 
Namely, 

\m 

\n 1) $n$-ary -- i.e.\ $n$-slot -- brackets that are totally antisymmetric and 
                                                     obey a fundamental identity; 
generalizing the $n = 2$ -- binary -- case's antisymmetric Lie bracket obeying Jacobi's identity.   
The $n = 3$ -- ternary -- case is then minimal for `nontrivial Nambuness', 
by which it is often the particular case considered \cite{Nambu}, including in the current Article. 

\m 

\n 2) In the classical canonical setting, letting one's bracket additionally be a derivation gives the Poisson bracket in the Lie case, 
and a Nambu--Poisson bracket in the Nambu case.
This corresponds to models with multiple Hamiltonians.  

\m 

\n 3) There are some difficulties with odd $n$-arity, but $n$-arity can always be increased, 
in particular permitting passage from $(2 \, p - 1)$-ary to $2 \, p$-ary brackets.
Because of this, there are some more detailed senses in which quaternary Nambu brackets are minimal.
Even $n$-arity can moreover be resolved as a sum of products of binary brackets.

\m 

\n 4) Nambu and Nambu--Poisson algebraic structures are $n$-ary counterparts of Lie and Poisson algebraic structures.
Our term `algebraic structure' means a portmanteau of algebra and algebroid cases.

\m

\n The current Article's main focus is on {\it Nambu Constraint Closure} in Sec \ref{NACC}, 
{\it Nambu Spacetime Generator Closure} in Sec \ref{NASGC}, 
and canonical and spacetime Nambu observables in Secs \ref{NO} and \ref{NSO} respectively. 
Sec \ref{NASGC} involves a       {\it Nambu Algorithm} analogue of the Lie   Algorithm \cite{Lie, Higher-Lie, XIV}, and 
Sec \ref{NACC} a somewhat more extensive {\it Nambu--Dirac Algorithm} analogue of the Dirac Algorithm \cite{Dirac, HTBook, ABook, III}
Both result in Nambu algebraic structures.
We consider in particular how first-class and second-class and Dirac-type brackets are manifested in Nambu Theory.  
We subsequently define notions of observables for Nambu Theory, in both in the spacetime and in the canonical settings.   
On the one hand, we eliminate various a priori possibilities \cite{AObs} by showing that multiple uses of the fundamental identity causes these to coincide.   
On the other hand, the remaining notions of Nambu observables form a lattice of {\it Nambu observables subalgebras}   
                                                                dual to that of {\it Nambu constraint subalgebraic structures}.

\m 

\n The particular focus above follows from ALToBI splitting into five super-aspects, as follows. 

\m 

\n{\bf Super-aspect i)}   Relationalism \cite{I, II, III, V, VI, X}. 

\m 

\n{\bf Super-aspect ii)}  Constraint, or Generator, Closure \cite{III, X}. 

\m 

\n{\bf Super-aspect iii)} Assignment of Observables \cite{III, VIII}.

\m 

\n{\bf Super-aspect iv)}  Construction (e.g. Spacetime Construction from space, or Construction of more structure from less assumed) \cite{Higher-Lie, III, IV, IX}. 

\m 

\n{\bf Super-aspect v)}   Reallocation of Intermediary Object (RIO) Invariance: a generalization of GR's Refoliation Invariance \cite{Higher-Lie, III, XII}  

\m 

\n These are furthermore the same i) to v) labels as used in our Abstract. 
By this, on the one hand, one can see that standard Lie derivatives remaining applicable in Nambu Mathematics 
has the consequence that the Relationalism that these implement remains rather similar to the standard Lie case's, lessening our need to cover i).  
On the other hand, iv) and v) have been argued \cite{Higher-Lie} to be more mathematically advanced \cite{Higher-Lie, XIV}, by which less is known about these.  
This leaves us with ii) and iii) as the main areas of interest, though with iv) and v) also having been argued to be selection principles 
within the Comparative Theory of Background Independence, posing the Nambu versions of these is briefly entertained in Sec 7.  

\m 

\n Various more general types of brackets \cite{S40-53-N55, Gengoux, V90, KS} are outlined in the Conclusion; these are of interest as regards 
Deformation Quantization \cite{L78, S98, Gengoux, Kontsevich} and Quantum Operator Algebras \cite{KRBook, Landsman}. 

\m 

\n As some further motivation, the Conclusion also explains how ternary Nambu brackets feature in 
Bagger--Lambert \cite{BL, BL2} and Gustavsson \cite{Gustavsson}'s (BLG) approach to M-Theory, at the level of brane worldsheet actions' potential terms. 
Entertaining this example furthermore justifies our inclusion of not only GR examples but also Supergravity ones: the low-energy limit of M-theory. 
We find moreover that all three of GR, Supergravity and BLG differ qualitatively as regards realization of Background Independence.

\section{Nambu Mathematics}\label{NM}

\subsection{Nambu brackets}

\n Let $\Frm$ be a real manifold and $\Frc^{\infty}(\Frm)$ the smooth $\mathbb{R}$-valued functions thereover.

\m 
	
\n{\bf Definition 1} The {\it Nambu bracket} alias $n$-{\it Lie bracket} is an $n$-ary $\mathbb{R}$-multilinear operation on $\Frc^{\infty}(\Frm)$, 
\be 
\ll  \m  \cb  ... \,  \cb  \m  \rl                                                                                                       \mmc 
\Frc^{\infty}(\Frm)   \times   \Frc^{\infty}(\Frm)   \times ... \,  \times  \Frc^{\infty}(\Frm)  \mlongrightarrow  \Frc^{\infty}(\Frm)  \m  
\ee
with the following properties. 

\m 

\n 1) It is totally antisymmetric: 
\be 
\ll  F_{i_1}   \cb  ... \,  \cb  F_{i_n}   \rl  \es 
\ll  F_{[i_1}  \cb  ... \,  \cb  F_{i_n]}  \rl                                                  \m . 
\label{anti}
\ee   
\n 2) It obeys the {\it fundamental identity} (sometimes alias {\it Filippov identity} \cite{Filippov}),  
\be
        \ll  F_1       \cb ... \,  \cb  F_{n - 1}    \cb 
        \ll  G_1       \cb ... \,  \cb  G_{n}        \rl       \rl                       \es  
\sumkn  \ll  G_1       \cb ... \,  \cb  G_{k - 1} 
        \ll  F_1       \cb ... \,  \cb  F_{n - 1}    \cb  G_k  \rl
			 G_{k + 1} \cb ... \,  \cb  G_n                    \rl                       \m .
\label{FI}
\ee
\n{\bf Example 1} For $n = 1$, antisymmetry supports only the zero function, so the `unary Nambu bracket' is trivial.

\m 
 
\n{\bf Example 2} For $n = 2$, the corresponding `binary Nambu bracket' is just the usual {\it Lie bracket}  
\be 
\ll  \m  \cb  \m  \rl                                                                           \mmc 
\Frc^{\infty}(\Frm)  \times  \Frc^{\infty}(\Frm)  \mlongrightarrow  \Frc^{\infty}(\Frm)        \m .
\ee
This is antisymmetric 
\be 
\ll  F  \cb  G  \rl  \es  - \ll  G  \cb  F  \rl                                                 \m , 
\ee
whereas (\ref{FI}) collapses to 
\be 
\ll       E  \cb  \ll  F       \cb  G  \rl \rl  \es  
\ll  \ll  E  \cb       F  \rl  \cb  G  \rl      \mmp                                         
\ll       F  \cb  \ll  E       \cb  G  \rl \rl  \m , 
\ee 
which can readily be identified as the {\it Jacobi identity}, 
\be 
\ll  E  \cb  \ll  F  \cb  G  \rl  \rl  \mmp  \mbox{cycles}  \es  0  \m .
\ee  
\n{\bf Example 3} For $n = 3$, the corresponding {\it ternary Nambu bracket} is thus the minimal nontrivially-Nambu bracket, 
\be 
\ll  \m  \cb  \m  \cb  \m  \rl                                                                                        \mmc 
\Frc^{\infty}(\Frm)  \times  \Frc^{\infty}(\Frm)  \times  \Frc^{\infty}(\Frm)  \mlongrightarrow  \Frc^{\infty}(\Frm)  \m . 
\ee							
This obeys the obvious total antisymmetry and the fundamental identity
\be 
\ll       A  \cb  B  \, \ll  C  \cb  D  \cb  E  \rl  \rl   \es  
\ll  \ll  A  \cb  B  \cb     C  \rl  D  \cb  E  \rl                \mmp 
\ll       C  \cb        \ll  A  \cb  B  \cb  D  \rl     \cb  E  \rl        \mmp
\ll       C  \cb  D  \, \ll  A  \cb  B  \cb  E  \rl     \rl                \m .
\ee	
\n{\bf Example 4} We shall see later that even-$n$ and odd-$n$ Nambu brackets differ in behaviour, 
including in a way that is relevant to the current Article's applications. 
Because of this, for some purposes $n = 4$'s {\it quaternary Nambu bracket} is the minimal nontrivially-Nambu example.  

\m 

\n{\bf Remark 1} One can furthermore entertain the {\it graded} (alias {\it supersymmetric}, or just {\it super}) versions of these brackets. 
The minus signs thus introduced do not however change any of the current Article's arguments, 
so this generalization is not explicitly exhibited, keeping presentation simpler.

\subsection{Nambu--Poisson brackets and Nambu Mechanics}\label{NMech}

\n{\bf Definition 1} The {\it Nambu--Poisson} bracket alias $n$-{\it Poisson bracket}
\be 
\lp  \m  \cb  ... \,  \cb  \m  \rp  \m 
\ee
is a Nambu bracket, thus obeying $\mathbb{R}$-multilinearity, total antisymmetry and the fundamental identity, 
but now also the {\it product rule} alias {\it Leibniz rule} alias {\it derivation property}, 
\be 
      \lp  F_1  \cb ... \,  \cb  F_{n - 1}  \cb  ... \,  \cb E \, G \rp        \es 
E \,  \lp  F_1  \cb ... \,  \cb  F_{n - 1}  \cb  ... \,  \cb G      \rp      \mmp 
      \lp  F_1  \cb ... \,  \cb  F_{n - 1}  \cb  ... \,  \cb  E      \rp\, G    \m . 
\label{deriv}
\ee 
\n{\bf Remark 1} It suffices to define this in just the one argument by total antisymmetry.  

\m 

\n{\bf Remark 2} Using this in Nambu Mechanics corresponds to configurations $\biQ$ being accompanied by multiple momenta $\u{\,\biP\,}$. 
%

\m 

\n{\bf Definition 2} The space of configurations and multiple momenta, as equipped by the Nambu--Poisson bracket, is termed generalized phase space. 
Let us denote this by 
\be 
\Phase(\bupSigma, \, n)  \es  \langle \, \biQ, \, \u{\,\biP \,} \,  \lp  \m  \cb  \m \,  ... \, \cb \m \rp \, \rangle   \m ,  
\ee 
where the $n$-label keeps track of the $n$-tuple of 1 configuration and $n - 1$ momenta, and of the corresponding $n$-slot Nambu--Poisson bracket. 

\m 

\n{\bf Example 2} For $n = 2$, the `binary Nambu--Poisson bracket' is just the usual Poison bracket, corresponding to there being just one momentum 
per configuration, by which the `binary generalized phase space' is just the usual phase space, 
\be 
\Phase(\bupSigma, \, 2)  \es  \Phase(\bupSigma)  \m . 
\ee 
\n{\bf Remark 3} It is thus the ternary Nambu--Poisson bracket that is minimally nontrivially-Nambu.  

\m 

\n{\bf Structure 1} Nambu Mechanics moreover has {\it multiple Hamiltonians}. 
In the current Article we view these as forming a {\it Nambu--Hamiltonian vector}, indicated by a wide underline, 
\be 
\NH(n)  \mma \mbox{ usually shortened to just } \m  \NH  \m . 
\ee  
In the $n$-ary case, this is an $(n - 1)$-vector. 
So for $n = 2$, one has the usual single Hamiltonian, whereas $n = 3$ is minimal for to have a vector of Hamiltonians, this case having a 2-vector thereof.  

\m 

\n{\bf Remark 4} The {\it Nambu--Hamilton equations of motion} are 
\be 
\frac{\d \biQ}{\d t}          \es  \lp  \biQ         \cb  H_1  \cb ... \,  \cb  H_{n - 1}  \rp  \m , 
\ee
\be 
\frac{\d \u{\,\biP\,}}{\d t}  \es  \lp  \u{\,\biP\,} \cb  H_1  \cb ... \,  \cb  H_{n - 1}  \rp  \m .   
\ee
For $n = 2$, these of course collapse to just the usual Poisson form of Hamilton's equations, 
\be 
\frac{\d \biQ}{\d t}  \es  \lp  \biQ  \cb  H  \rp  \m ,
\ee 
\be 
\frac{\d \biP}{\d t}  \es  \lp  \biP  \cb  H  \rp  \m .
\ee 
For $n = 3$, the minimally nontrivally-Nambu ternary Nambu--Hamilton's equations are 
\be 
\frac{\d \biQ}{\d t}          \es  \lp  \biQ          \cb  H_1  \cb  H_2  \rp  \m ,
\ee 
\be 
\frac{\d \u{\,\biP\,}}{\d t}  \es  \lp  \u{\,\biP\,}  \cb  H_1  \cb  H_2  \rp  \m .
\ee 
\n{\it Nambu conservation} takes the form  
\be 
0  \es  \frac{\d F}{\d t}     \es  \lp  F  \cb  H_1  \cb ... \,  \cb  H_{n - 1}  \rp  \m . 
\ee 
%
%
For $n = 2$, this collapses to the standard Hamiltonian conservation equation,   
\be 
0  \es  \frac{\d F}{\d t}  \es  \lp  F  \cb  H              \rp  \m . 
\ee
For the minimal nontrivially-Nambu $n = 3$ ternary bracket case, 
\be 
0  \es  \frac{\d F}{\d t}  \es  \lp  F  \cb  H_1  \cb  H_2  \rp  \m . 
\ee

\subsection{Changing $\bin$-arity}\label{Change-n}

\n{\bf Structure 1} The $n$-ary Nambu--Poisson bracket can be expressed as the $n \times n$ Jacobian \cite{Nambu}
\be 
\lp F_{i_1}  \cb  ... \,  \cb  F_{i_n}  \rp                                                                                  \es  
\left|  \frac{  \pa(  F_{i_1}, \, ... \, , \,  F_{i_n}  )  }{  \pa(  \biQ,  \,  \biP_{i_1}, \, ... \,  \biP_{i_{n - 1}}  )  }  \right|  \m .  
\label{Jacobian}
\ee 
%
%
\n{\bf Structure 2} $(2 \, p - 1)$-ary brackets can be locally embedded into $2 \, p$-ary ones according to \cite{odd-even, Nambu-Rev} 
\be 
\lp  x_a  \cb  F_1  \cb  ... \,  \cb  F_{2 \, p - 1}  \rp                                                                                   \es  
\left|  \frac{  \pa(  x_a, \,  F_1, \,  ... \, , F_{2 \, p - 1}  )                  }{  \pa(  x_1, \,  ... \, , x_{2 \, p - 1}  )  }  \right|  \es  
\left|  \frac{  \pa(  x_a, \,  F_1, \, ...  \, , F_{2 \, p - 1} , \, x_{2 \, p}  )  }{  \pa(  x_1, \,  ... \, , x_{2 \, p}      )  }  \right|  \es  
\lp  x_a  \cb  F_1  \cb  F_2     \cb  x_{2 \, p}      \rp                                                                                   \m .   					 
\ee
for $a$ running over 1 to $2 \, p - 1$.
%

\m 

\n{\bf Structure 3} $2 \, p$-ary brackets can be resolved to a totally antisymmetric sum,  
\be 
                                                     \lp  I_{i_1}            \cb  I_{i_2}         \cb  ... \,  \cb  I_{i_{2\,p}}  \rp  \es  
\frac{1}{  2^p p!  }  \epsilon^{i_1 ... i_{2 \, p}}  \lp  I_{i_1}            \cb  I_{i_2}         \rp  
                                                     \lp  I_{i_3}            \cb  I_{i_4}         \rp  \, ... \, 
									    	         \lp  I_{i_{2 \, p -1}}  \cb  I_{i_{2 \, p}}  \rp                                   \m .   
\ee 
{\bf Example 1} The minimal nontrivial example of this is the quaternary \cite{CZ02}
\be 
                                     \lp  I_1      \cb  I_2      \cb           I_3      \cb  I_4      \rp  \es  
\frac{1}{8}  \epsilon^{i_1 ... i_4}  \lp  I_{i_1}  \cb  I_{i_2}  \rp  \,  \lp  I_{i_3}  \cb  I_{i_4}  \rp  \es 
                                     \lp  I_1      \cb  I_2      \rp  \,  \lp  I_3      \cb  I_4      \rp  \mmm 
                                     \lp  I_1      \cb  I_3      \rp  \,  \lp  I_2      \cb  I_4      \rp  \mmm 
                                     \lp  I_1      \cb  I_4      \rp  \,  \lp  I_2      \cb  I_3      \rp  \m . 
\ee 
In the current Article, however, 
we complement the literature by directly setting up physical structure using $n$-ary brackets rather than resolving down to binary brackets.

\subsection{Nambu and Nambu--Poisson algebras and groups}

\n{\bf Definition 1} A {\it Nambu} alias $n$-{\it Lie} (occasionally alias {\it Filippov algebra})  
is a vector space $\FrV$ equipped with an internal $n$-ary bracket satisfying (\ref{anti}, \ref{FI}) and such that
\be 
\rl  \uc{\sbcg}  \cb ... \,  \cb  \uc{\sbcg}  \rl  \es  \s{\uc{\biG}}{\uc{...}} \, \uc{\sbcg}  \m .   
\ee
The `...' here  indicates underlining enough for an $(n + 1)$-array of constants $\biG$.
\n{\bf Definition 2} If \ref{deriv}) holds as well, we have a {\it Nambu--Poisson} algebra (some authors use Nambu or $n$-Lie in this case as well).   

\m 

\n For $n = 2$, 
\be 
\lp  \uc{\sbcg}  \cb  \uc{\sbcg}  \mbox{\bf \}}  \es  \uc{\uc{\uc{\biG}}} \, \uc{\sbcg}             \m ,  
\ee 
so the $\biG$ form the standard 3-array of {\it structure constants} of a Lie algebra; this is moreover a Poisson algebra because of the derivation property.     

\m 

\n For the minimally nontrivially-Nambu ternary case,
\be 
\lp  \uc{\sbcg} \cb  \uc{\sbcg}  \cb  \uc{\sbcg}  \rp  \es  \uc{\uc{\uc{\uc{\biG}}}} \, \uc{\sbcg}  \m ,  
\ee 
for $\biG$ a 4-array of Nambu structure constants. 

\m 

\n{\bf The current Article's Notation}  
We use coordinate-free expressions as much as possible, with $\uc{\m \m}$ to distinguish generator or constraint-valued vectors and arrays 
from space or spacetime valued ones (underlines and overarrows) and $\uo{\m \m}$ for observables valued ones.  
Small calligraphic font denotes generators -- lower case -- and constraints -- upper case -- 
with the associated observables being matchingly upper and lower case typeface font. 
When required, generators carry $\fg$ indices, constraints $\fC$ indices, spacetime observables $\fo$ indices and canonical observables $\fO$ indices.   
These conventions allow one to pick out notions of each kind in our formulae, 
which is important in successfully navigating the Problem of Time and Background Independence. 
For instance, a detailed treatise might require over 30 types of objects each carrying a different type of index, 
so it is useful for, firstly, its 12 types of observables, say, to come in the same font, further partitioned by case into spacetime and canonical versions. 
Secondly, for the 12 different types of observables indices to themselves have a distinctive matching font, 
{\sl and} be as absent as possible by use of coordinate-free notation.   
A further useful convention is to distinguish between objects and the spaces those objects form, such as between configurations and configuration spaces, 
generators and algebraic structures of generators, or observables and algebras of observables. 
We attain this by using the bold mathfrak font for the leading letter of each space of objects.
We help the reader keep track of these by using -$\bFrg$en, -$\bFrC$on, -$\bFro$bs and $\bFrO$bs as endings for the symbols for 
spaces of generators, constraints, observables associated with generators, and observables associated specifically with constraints respectively.  

\m 
 
\n{\bf Remark 1} Nambu groups and Nambu--Poisson groups are global counterparts for the above; 
on the one hand, this requires a somewhat nontrivial definition for its multiplication operation \cite{V98}; 
on the other hand, the current Article's local pursuits preclude further elaboration.

\subsection{From Lie algebras to Lie algebroids}

\n{\bf Remark 1} All we need to know about Lie algebroids for the current work is that now 
\be 
\lp  \uc{\sbcg}  \cb  \uc{\sbcg}  \rp  \es  \uc{\uc{\uc{\biG}}}(\biB) \, \uc{\sbcg}  \m ,  
\ee 
where the $\biG(\biB)$ are now a 3-array of {\it structure functions}, rather than a Lie algebra's structure constants.   
Their functional dependence is on some base objects (e.g.\ configuration or phase space variables).  
 
\m

\n Some general motivation for extension form Lie algebras to Lie algebroids is as follows. 

\m 

\n{\bf Motivation 1} Robustness of Algorithms (Sec 3) as regards the classical Deformation Theory used for Construction (Sec 7) leads to Lie algebroids arising.  

\m 

\n{\bf Motivation 2} The canonical approach to GR necessitates an algebroid: the {\it Dirac algebroid} \cite{Dirac, T73, BojoBook}  
\be
\lp  ( \, \u{\bscM}  \,  |  \,  \u{\bupxi}  \, ) \cb  ( \,  \u{\bscM}  \,  |  \,  \u{\bupchi}  \, )  \rp  \es  (  \u{\bscM}             \, | \,  \u{[ \, \bupxi, \, \bupchi \, ]}  )  \m ,
\label{Mom,Mom}
\ee
\be
\lp  ( \, \scH      \,  |  \,  \upzeta        \, ) \cb  ( \,  \u{\bscM}  \,  |  \,  \u{\bupxi}   \, )  \rp  \es  ( - \Lie_{\sbupxi} \scH  \, | \,  \upzeta  )                              \m , 
\label{Ham,Mom}
\ee
\be 
\lp  ( \, \scH     \,  |  \,  \upzeta       \, ) \cb  ( \,  \scH       \,  |  \,  \upomega     \, )  \rp  \es   ( \, \u{\bscM} \cdot \u{\u{\bh}}^{-1} \cdot  \, | \,  \upzeta \, \overleftrightarrow{\u{\bpa}}  \upomega  \, ) \m . 
\label{Ham,Ham}
\ee
$( \m | \m )$ is here the integral-over-space functional inner product, 
$[ \m , \m ]$, the differential-geometric commutator Lie bracket, 
$\bh^{-1}$ is the inverse of the spatial metric, 
and $\bupxi$, $\bupchi$, $\zeta$ and $\upomega$ are smearing functions. 
Such `multiplication by a test function' serves to render rigourous a wider range of 
     `distributional' manipulations \cite{AMP} provided that these occur under an integral sign.  

\m 

\n In particular, the third bracket involves not structure constants but structure functions, out of containing  $\bh^{-1}(\bh(\underline{x}))$ 

\m 

\n Canonical Supergravity moreover requires an algebroid that is qualitatively distinct from this (see Sec 3.12).  

\m 

\n{\bf Motivation 3} Kinematical quantization's \cite{M63, I84} modern reformulation \cite{Landsman} in terms of Lie algebroids. 

\m 

\n{\bf Remark 1} We thus introduce the portmanteau {\sl Lie algebraic structure} to jointly encompass Lie algebras and Lie algebroids; 
{\it constraint algebraic structures} are meant in this sense.  

\m 

\n{\bf Structure 2} Much as topologicial and other global considerations require Lie groups rather than Lie algebras, 
they require Lie groupoids \cite{Landsman} instead of Lie algebroids.

\subsection{Corresponding extension to Nambu(--Poisson) algebroids}

\n{\bf Remark 1} Suitable extensions of this motivation apply in the Nambu case as well, by which Nambu algebroids and groupoids \cite{V98, W02} enter contention.  
One would moreover expect an M-Theory of comparable-to-greater complexity to GR, Supergravity... to retain algebroid character in its constraint algebraic structure,  

\m 

\n{\bf Definition 1}  
\be 
\lp  \uc{\sbcg}  \cb  ... \,  \cb  \uc{\sbcg}  \rp  \es  \s{\uc{\biG}}{\uc{\mbox{\bf...}}}(\biB) \, \uc{\sbcg}  \m ,  
\ee 
for $\biG$ now an $(n + 1)$-array of Nambu structure functions, and generalized phase space variables being a particularly relevant chose of basic objects.

\m

\n{\bf Example 1} Clearly for $n = 2$, this collapses to the preceding subsection's Lie algebroid. 

\m  

\n{\bf Example 2} For $n = 3$'s minimally nontrivially-Nambu ternary algebroid, 
\be 
\lp  \uc{\sbcg}  \cb  \uc{\sbcg}  \cb  \uc{\sbcg}  \rp  \es  \uc{\uc{\uc{\uc{\biG}}}}(\biB) \, \uc{\sbcg}  \m , 
\ee 
for $\biG(\biB)$ now a 4-array of Nambu structure functions. 

\m 

\n{\bf Remark 1 2} Nambu groupoids, on the other hand, would appear very largely not to have been studied to date.

\section{The Nambu--Dirac Algorithm for Constraint Closure}\label{NACC}

\n Given an initial set of constraints, whether from `thin air' or from relational constraint providers, 
{\it Nambu--Poisson brackets} are introduced to assess Closure via the below {\it Nambu--Dirac Algorithm}.

\subsection{Notions of equality}\label{NoE}

The following carries over from the Dirac \cite{Dirac} case, so it is just a matter of introduction of notation for the current Article.  

\m 

\n{\bf Definition 1} Dirac's \cite{Dirac} notion of  
\be 
\mbox{\it weak equality} \mapprox  
\ee 
means equality up to additive functionals of the constraints.
In contrast, `strong equality' = just means equality in the usual sense.  
The current Series also uses 
\be 
\mbox{`{\it portmanteau equality}' } \peq  
\ee 
to encompass both strong and weak equality; this is useful in more summarily introducing notions that have strong and weak versions.

\subsection{Nambu primary and Nambu secondary constraints}\label{NPS}

\n{\bf Definition 1} In the standard Hamiltonian formulation, {\it constraints} are taken to be relations 
\beq 
\sbcC(\biQ, \biP) = 0
\eeq 
between the momenta $\biP$ by which these are not independent.
This is quite a general type of constraint considered by Dirac \cite{Dirac}, suitable for much of classical Fundamental Physics; 
see e.g.\ \cite{Lanczos, ABook} for discussion of furtherly general notions of constraint. 

\m 

\n{\bf Definition 2} In the Nambu--Hamiltonian vector formulation of Nambu, {\it constraints} are relations 
\beq 
\sbcC(\biQ, \u{\,\biP\,}) = 0
\eeq 
between the multiple momenta $\u{\,\biP\,}$ by which these are not independent.

\m 

\n{\bf Definition 2} Constraints arising from the form of the Lagrangian via non-invertibility of momentum--velocity relations alone are termed {\it primary} 
\cite{Dirac, HTBook}. 
We denote these by 
\be 
\bscP \mma \mbox{ indexed by } \m  \fP   \m . 
\ee     
\n{\bf Definition 3} Constraints furthermore requiring input from the variational equations of motion are termed {\it secondary} \cite{Dirac, HTBook}. 
We denote these by 
\be 
\bscS \mma \mbox{ indexed by } \m  \fS   \m . 
\ee 

\m 

\n{\bf Remark 2} In the usual Hamiltonian case corresponding to one momentum per configuration and Poisson brackets.

\subsection{Two first instances of Nambu--Dirac multiplier-appending of constraints}\label{NDMA-1}

\n{\bf Remark 1} One part of Dirac's approach \cite{Dirac} to handling constraints involves appending them additively with Lagrange multipliers 
to a system's incipient or `bare' Hamiltonian, $H$ (see Sec \ref{ND-Brackets} for the other part).  
We extend this here to Nambu--Hamiltonian vectors, for which the Lagrange multipliers require a matching vectorial index.  

\m 

\n{\bf Definition 1} The {\it Nambu--Dirac `starred' Hamiltonian} \cite{Dirac} is the result of appending a formalism of a theory's primary constraints $\bscP$ 
with a priori arbitrary phase space functions $\biA(\biQ, \u{\,\biP\,})$ in the role of Lagrange multipliers, 
\be 
\NH^*      \:=  \NH  \mmp  \u{\,\uc{\biA}\,} \cdot \uc{\bscP}  \m . 
\ee 
\n{\bf Definition 2} The {\it Nambu--Dirac total Hamiltonian} \cite{Dirac} 
is the result of appending such instead with {\it unknown functions} $\biu(\biQ, \u{\,\biP\,})$ 
\be 
\NH_{\sT}  \:=  \NH  \mmp  \u{\,\uc{\biu}\,} \cdot \uc{\bscP}   \m .
\ee 
\n{\bf Remark 2} The intent in the unknown case is to {\sl solve for} some multipliers. 

\m 

\n 1) In Dirac's case, binarity renders this a linear and thus readily solvable equation in $\biu$ \cite{Dirac}: 
\beq
0  \mapprox   \dot{\bscP}    
     \es      \lp  \bscP  \cb       H      \rp                     \mmp  
          	  \lp  \bscP  \cb  \uc{\bscP}  \rp \cdot {\uc{\biu}}   \m .
\label{de-for-u}
\eeq
\n 2) In nontrivially-Nambu cases, however, one faces the complication of a nonlinear equation in $\u{\,\biu\,}$. 
E.g.\ for $n = 3$'s ternary Nambu brackets case, 
\beq
0  \mapprox  \dot{\bscP}    
      \es    \lp  \bscP  \cb     H_1        \cb      H_2                \rp                                                \mmp 
          	 \lp  \bscP  \cb     H_1        \cb  \uc{\bscP}             \rp  \cdot  \uc{\biu}\mbox{}_1                     \mmp 
			 \lp  \bscP  \cb  \uc{\bscP}    \cb      H_2                \rp  \cdot  \uc{\biu}\mbox{}_2                     \mmp 
			 \lp  \bscP  \cb  \bscP_{\sfP}  \cb  \bscP_{\sfP^{\prime}}  \rp  \,     \biu_{\sfP \, 1}\biu_{\sfP^{\prime}2}  \m .
\label{nonlinear-de-for-u}
\eeq

\subsection{Nambu--Dirac's Little Algorithm}\label{NDLA}

\n{\bf Definition 1} {\it `Nambu--Dirac's Little Algorithm'}\footnote{This is named in honour of Nambu, 
rather than by Nambu, and is a direct parallel of Chapter 1 of \cite{Dirac}.} 
This consists of evaluating Nambu--Poisson brackets between a given input set of constraints so as to determine whether these are consistent and complete.  
Four possible types of outcome are allowed in this setting.    

\m 

\n{\bf Type 0)} {\it Inconsistencies}.
That this is possible in the Principles of Dynamics is clear from e.g.\ the Lagrangian 
\be 
L = \dot{q} + q                     \m ,
\ee 
in standard (binary) Poisson brackets context, giving as its Euler--Lagrange equations  
\be 
0 = 1                               \m . 
\ee
Dirac's envisaging the need to allow for the possibility of inconsistency allows for algorithms with {\sl selection principle} properties.  

\m 

\n{\bf Type 1)} {\it Mere identities} -- equations that reduce to 
\be 
0 \peq 0                       \m ,
\ee 
comprising strong identity 
\be 
0 = 0
\ee 
and the more general weak identity 
\be 
0 \approx 0                             \m.  
\ee 
\n{\bf Type 2)} {\it Further secondary constraints}, i.e.\  equations independent of the Lagrange multiplier unknowns. 
Arising in this (implicitly variational) manner lies within the remit of secondary, rather than primary, constraints.

\m 

\n{\bf Type 3)} Relations amongst some of the appending Lagrange multipliers functions themselves. 
These are not constraints but a further type of equation, termed `{\it specifier equations}' since they specify restrictions on the Lagrange multipliers; 
see \cite{VII} for more about these.

\subsection{Nambu--Dirac Algorithm termination conditions}\label{NDATC}

\n{\bf Remark 1} If type 0) occurs, the candidate theory is inconsistent. 

\m 

\n{\bf Definition 1} Let us refer to equations of all types arising from Nambu--Dirac Algorithms bar 0) as {\it ab initio consistent}.

\m 

\n{\bf Definition 2} With type 1)'s mere identities having no new content, let us call types 2) to 4) `{\it nontrivial ab initio consistent objects}'.  
Note that we say `objects', not `constraints', to include type 3)'s specifier equations.  

\m 

\n{\bf Remark 2} If type 2) occurs, the resultant constraints are fed into the next iteration of the Nambu--Dirac Algorithm, 
which starts with the extended set of objects.  
One is to proceed thus recursively until one of the following termination conditions is attained. 

\m  

\n{\bf Termination Condition 0} {\it Immediate inconsistency} due to at least one inconsistent equation arising.

\m 

\n{\bf Termination Condition 1} {\it Combinatorially critical cascade}. 
This is due to the iterations of the Nambu--Dirac Algorithm producing a cascade of new objects 
down to the `point on the surface of the bottom pool' that leaves the candidate with no degrees of freedom.   
I.e.\ a combinatorial triviality condition.    

\m 

\n{\bf Termination Condition 2} {\it Sufficient cascade}, i.e.\ running `past the surface of the bottom pool' of no degrees of freedom 
into the `depths of inconsistency underneath'.

\m 

\n{\bf Termination Condition 3} {\it Completion} is that the latest iteration of the Nambu--Dirac Algorithm has produced no new nontrivial consistent equations, 
indicating that all of these have been found. 

\m 

\n{\bf Remark 3} Our input candidate set of constraints is either itself {\it complete} 
                                                     or {\it incomplete} -- `nontrivially Nambu--Dirac' -- 
													 depending on whether it does not or does imply any further nontrivial objects.
If it is incomplete, it may happen that the Nambu--Dirac Algorithm provides a completion, by only an {\it combinatorially insufficient cascade} arising, 
from the point of view of killing off the candidate theory.  
														 
\m 														 

\n{\bf Remark 4} So, as the first of three options, Termination Condition 3) is a matter of acceptance 
of an initial candidate set of constraints alongside the cascade of further objects emanating from it by the Nambu--Dirac Algorithm.   
(This acceptance is the point of view of consistent closure; further selection criteria might apply.)    			
This amounts to succeeding in finding -- and demonstrating -- a `Nambu--Dirac completion' of the incipient candidate set of constraints.  											 

\m  

\n{\bf Remark 5} As the second of three options, Termination Conditions 0) and 2) are a matter of rejection of an initial candidate set of constraints.  
The possibility of either of these applying at some subsequent iteration justifies our opening conception in terms of ab initio consistency.

\m 

\n{\bf Remark 6} As the third of three options, Termination Condition 1) is the critical case on the edge between acceptance and rejection; 
further modelling details may be needed to adjudicate this case.

\m 

\n{\bf Remark 7} In each case, {\it functional independence} \cite{Lie} is factored into the count made; 
the qualification `combinatorial' indicates Combinatorics not always sufficing in having a final say.  
For instance, Field Theory with no local degrees of freedom can still possess nontrivial global degrees of freedom. 
Or Relationalism can shift the actual critical count upward from zero, e.g.\ by requiring a minimum of two degrees of freedom 
so that one can be considered as a function of the other.  
If this is in play, we use the adjective `relational' in place of (or alongside) `combinatorial'.  

\m 

\n{\bf Remark 8} If type 2)'s further constraints occur, these are fed into the subsequent iteration of the Nambu--Dirac Algorithm.

\subsection{Nambu first- and second-class constraints}\label{NFSC}

\n This extends Dirac's well-known distinction in the binary case with Poisson brackets.

\m 

\n{\bf Definition 1} {\it Nambu-first-class constraints} (generalizing \cite{Dirac, HTBook})  
\be 
\bscF \m \mbox{ indexed by } \m \fF
\ee  
are those that close amongst themselves under Nambu--Poisson brackets.  

\m 

\n{\bf Definition 2} {\it Nambu-second-class constraints} \cite{Dirac, HTBook} 
\be 
\Sec \m \mbox{ indexed by } \m \fE
\ee 
are defined by exclusion to be those that are not Nambu-first-class. 

\m 

\n{\bf Remark 1} Our notation is subject to the letter `S' already being in use for Nambu-secondary constraints.    

\m 

\n{\bf Remark 2} In the ternary case, four kinds of Nambu--second-classness are a priori possible \cite{AObs}: blockwise splits 
\be 
\lp  \con  \cb  \con    \cb \bscD_1  \rp  \m \npeq \m  0  \m , 
\ee
\be 
\lp  \con  \cb  \bscD_2  \cb \bscD_2  \rp  \m \npeq \m  0  \m , 
\ee
\be 
\lp  \con  \cb  \bscD_1  \cb \bscT_1  \rp  \m \npeq \m  0  \m ,    
\ee
\be 
\lp  \con  \cb  \bscD_2  \cb \bscT_2  \rp  \m \npeq \m  0  \m ,    
\ee
However, the first pair are conceptually the same by change of block relabelling, rendering the second pair equal as well. 
Thus we are down to two notions.  

\m 

\n{\bf Nambu Class Non-Proliferation Theorem} First and second class distinction suffices for Nambu Theory. 

\m 

\n\Proof By definitions by exclusion, formulate our putative distinction in terms of multiple kinds of first-classness.  
This is then the same shape as the second and third parts of Sec 5's Nambu Observables Non-Proliferation Theorem. $\Box$

\subsection{Two further instances of Nambu--Dirac multiplier-appending of constraints}\label{NDMA-2}

\n{\bf Definition 1} The {\it Nambu--Dirac primed Hamiltonian vector} is 
\be 
\NH^{\prime}  \:=  \NH  \mmp  \u{\,\uc{\bip}\,} \cdot \uc{\bscP}                                   \m ,
\ee  
for which we provide the true name {\it particular-primary Hamiltonian}, signifying `with particular-solution multipliers appending primary constraints'.  

\m 

\n{\bf Definition 2} Finally, the {\it Nambu--Dirac extended Hamiltonian vector} is 
\be 
\NH_{\sE}  \:=  \NH   \mmp  \u{\,\uc{\biu}\,} \cdot \,\uc{\bscP}  \mmp  \u{\,\uc{\bia}\,} \cdot \uc{\bscS}     \m , 
\ee 
for arbitrary functions $\bia$ and specifically Nambu-first-class-secondary constraints $\bscS$.

\m 

\n{\bf Remark 1} Such a notion could clearly be declared for each iteration of the Nambu--Dirac Algorithm, 
with the above one coinciding with Nambu's Little Algorithm attaining completeness. 
In this sense, $\NH_{\sE}$ is a candidate theory's Nambu--Hamiltonian vector that is {\it maximally} extended by appending of Nambu-first-class constraints.

\subsection{Removing Nambu-second-class constraints}\label{ND-Brackets}

\n We are now to envisage the possibility of second-class constraints arising at some iteration in the Nambu--Dirac Algorithm.

\m 

\n {\bf Motivation} Removal of second-class constraints 
is especially relevant since many standard quantum procedures are based on just first-class constraints remaining by that stage.
This usually entails classical removal of any other nontrivial consistent entities which feature in the original formulation.

\m 

\n{\bf Remark 1} Second-class constraints can moreover be slippery to pin down. 
This is because second-classness is not invariant under taking linear combinations of constraints. 
Linear Algebra dictates that the invariant concept is, rather, {\sl irreducibly second-class constraints} \cite{Dirac, HTBook} 
\be 
\bscI \m \mbox{ indexed by } \m \fI  \m .
\ee  
\n{\bf Structure 1} In the $n = 2$ case, irreducibly second-class constraints can be factored in 
by redefining the incipent Poisson bracket with the  {\it Dirac bracket} 
\beq
\lp  F  \cb  G  \rd  \:=  \lp  F  \cb G           \rp                                                                                           \mmm  
                          \lp  F  \cb \uc{\bscI}  \rp  \cdot  \lp  \uc{\bscI}  \cb  \uc{\bscI}  \rp^{-1}  \cdot  \lp  \uc{\bscI}  \cb  G  \rp   \m . 
\eeq
Here the --1 denotes the inverse of the given matrix, and each $\cdot$ contracts the underlined objects immediately adjacent to it.   
This formula amounts to projecting out the $\bscI$, by which they are rendered mere Casimirs with respect to the new bracket.  

\m

\n{\bf Structure 2} Even-$n$ Nambu--Poisson brackets \cite{GM98} admit generalizations. 
The most obvious reason for this restriction is that odd-$n$ arrays do not possess a natural notion of inverse. 

\m 

\n The first nontrivially-Nambu `Nambu--Dirac bracket' is thus the quaternary 
\beq 
\lp     A                         \cb      B                              \cb     C                         \cb     D                               \rd       \:=  
\lp     A                         \cb      B                              \cb     C                         \cb     D                               \rp       \mmm 
\lp     A                         \cb      B                              \cb  \bscI_{\sfI}                 \cb  \bscI_{\sfI^{\prime}}                        \rp 
\lp  \bscI_{\sfI}                 \cb  \bscI_{\sfI^{\prime}}              \cb  \bscI_{\sfI^{\prime\prime}}  \cb  \bscI_{\sfI^{\prime\prime\prime}}  \rp^{-1}  
\lp  \bscI_{\sfI^{\prime\prime}}  \cb  \bscI_{\sfI^{\prime\prime\prime}}  \cb     C                         \cb     D                               \rp       \m .  
\hspace{1in}  
\eeq
\n For the general $2 \, p$-ary case,   
$$
\lp  A_1            \cb  A_2     \cb  ... \,          \cb  A_{2 \, p}  \rd                         \:=  
\lp  A_1            \cb  A_2     \cb  ... \,          \cb  A_{2 \, p}  \rp                         \hspace{2in} 
$$
\be
\hspace{2in} \mmm 
\lp  A_1            \cb  ... \,  \cb  A_p             \cb  \bscI_1     \cb  ... \, \cb \bscI_p     \rp 
\lp  \bscI_1        \cb  ... \,  \cb  \bscI_{2 \, p}                                               \rp^{-1} 
\lp  \bscI_{p + 1}  \cb  ... \,  \cb  \bscI_{2 \, p}  \cb  A_{p + 1}   \cb  ... \, \cb A_{2 \, p}  \rp        \m .  
\ee
For odd-$n$, one can locally embed $(2 \, p - 1)$-ary brackets in $2 \, p$-ary brackets as per Sec (\ref{Change-n}), 
and then set up the corresponding even case's Nambu--Dirac brackets.  

\m 

\n We can also make use of $2 \, p$-ary brackets' re-expressability as products of binary brackets. 
 
\m 
 
\n{\bf Remark 2} The role of classical brackets role initially played by the Nambu--Poisson brackets may be taken over by the Nambu--Dirac brackets.

\m  

\n{\bf Structure 3} Use the extended Nambu-phase-space method counterpart of e.g.\ \cite{HTBook}'s account for the binary case's Poisson brackets.   

\m 

\n{\bf Remark 3} This approach is systematically available without restriction on $n$. 
I.e.\ this is Nambu-robust without having to transcend $n$-arity.

\subsection{Full Nambu--Dirac Algorithm}\label{NDA-Full}

Proceed as before, except that whenever second-class constraints appear, one switches to (new) Nambu--Dirac brackets that factor these in.  
This amounts to a fifth type of equation being possible, as follows.  

\m 

\n{\bf Type 4)} {\it Further Nambu-second-classness} can arise.

\m 

\n 4.i) This could be {\it Nambu-self-second-classness}, by which brackets between some new constraints do not close.  

\m 

\n 4.ii) This could instead be {\it Nambu-mutual-second-classness}, 
meaning that some bracket between a new constraint and a previously found constraint does not close. 
By which, this previously found constraint was just {\it hitherto Nambu-first-class}.
I.e.\ Nambu-first-classness of a given constraint can be lost whenever a new constraint is discovered.  

\m 

\n{\bf Remark 1} At each iteration, then, one ends up with a bare Nambu-Hamiltonian vector with Nambu-first-class constraints appended using multipliers.  
The final such is once again denoted by $\NH_{\sE}$, corresponding to having factored in all Nambu-second-class constraints 
and appended all Nambu-first-class constraints.  
Each other notion of Nambu-Hamiltonian vector above can also be redefined for Nambu-Dirac brackets.     

\m 

\n{\bf Remark 2} The $n = 2$ version of the above is as far as Dirac gets; subsequent discoveries in practice dictate the addition of the below sixth type. 
Dirac knew about this \cite{Dirac}, commenting on needing to be lucky to avoid this at the quantum level. 
But no counterpart of it enters his classical-level Algorithm.  

\m 

\n{\bf Type 5} {\it Discovery of a topological obstruction} to having a Nambu algebraic structure of constraints.
The most obvious examples of this are anomalies \cite{Bertlmann} at the quantum level in the binary case.  
It is however a general brackets phenomenon rather than specifically a quantum phenomenon, and admits an $n$-ary Nambu counterpart

\m

\n Two distinct strategies for dealing with this are as follows. 

\m 

\n{\bf Strategy 1} Set a cofactor of the topological term to zero when the modelling is permissible of this. 
In particular {\bf strongly vanishing} cofactors allow for this at the cost (or discovery) of fixing some of the theory's hitherto free parameters.

\m 

\n{\bf Strategy 2} Abandon ship. 

\m 

\n{\bf Remark 3} In the Nambu Little Algorithm, everything stated was under the aegis of all objects involved at any stage are Nambu-first-class, 
                                                                                         and that no topological obstruction terms occur.

\subsection{Nambu--Poisson constraint algebraic structures introduced}\label{NDCAS}

\n{\bf Structure 1} The end product of a successful candidate theory's passage through the Nambu--Dirac Algorithm 
is a Nambu--Poisson {\it constraint algebraic structure} consisting solely of first-class constraints closing under Nambu--Poisson (or more generally Nambu--Dirac) brackets. 

\m 

\n{\bf Modelling assumption} Assume that neither the topological nor tertiary complications explained below occur.

\m 

\n{\bf Structure 2} Under this condition, schematically, 
\beq
\lp  \bscF  \cb  \bscF  \rp  \peq  0                                                                     \m .
\label{F-F}
\eeq
This is a portmanteau for the strong version
\beq
\lp  \bscF  \cb  \bscF  \rp   \peq  0                                                                    \m , 
\label{F-F-S}
\eeq
and the weak version: 
\beq
\lp  \uc{\bscF}  \cb  \uc{\bscF}  \rp  \es  \uc{\uc{\uc{\biF}}} \cdot \uc{\bscF}                         \m .
\label{F-F-W}
\eeq
The $\biF$ can here be the structure constants of a Nambu algebra, or a Nambu algebroid's phase space functions, $\biF(\biQ, \biP)$.
\beq
\lp  \uc{\con}  \cb  \uc{\con}  \rp  \es  \uc{\uc{\uc{\biC}}} \cdot \uc{\con}  \mmp
                                            \uc{\uc{\uc{\biD}}} \cdot \uc{\bscD}  \mmp  \uc{\uc{\Theta}}  \m .
\label{F-F-W-2}
\eeq
as a more general output, with discovered constraints, $\bscD$. 
One then needs to look at $\lp \uc{\con} \, \cb \uc{\bscD} \rp$  
and                       $\lp \uc{\bscD} \, \cb \uc{\bscD} \rp$.  

\m 

\n In the $n$-ary case, 
\beq
\lp  \uc{\con}  \cb  ...  \, \cb  \uc{\con}  \rp  \es  \s{\uc{\biC}}  {\uc{\mbox{\bf...}}}  \cdot    \uc{\con}                                         \mmp
                                                                                                    \s{\uc{\biD}}{\uc{\mbox{\bf...}}}  \cdot  \uc{\bscD}  \mmp 
																								    \s{\uc{\bTheta}}{\uc{\mbox{\bf..}}}                   \m .      
\label{F-F-W-3}
\eeq
where the double dot indicates that $\bTheta$ is an $n$-array.

\subsection{Lattices from the Nambu--Dirac Algorithm}\label{ND-Latt}

\n{\bf Structure 1} There is further interest in finding the (in particular conceptually meaningful) 
consistent subalgebras supported by the top constraint algebra. 
The notions of constraint in question form a bounded lattice for each $n$,  
\be 
\lattice_{\scon(n)}  \m ,
\ee 
with the full repertoire of first-class constraints $\bscF$ as top element, and the absence of constraints $\emptyset$ as bottom element. 
We use $\bscF(n)$ as more specific notation for $n$-ary Nambu Theory's first-class constraints.
The other members of the lattice of notions of constraint are middle elements, the {\it notions of Z-algebraic structure} denoted by 
\be 
\bscZ \m \mbox{ with each type indexed by } \m  \fZ  \m .
\ee  
See row 2 of Fig \ref{C-Latt} for a schematic sketch.

\subsection{Constraint algebraic structures exemplified}

\n{\bf Structure 1}  Let us denote the {\it top space of classical Nambu-first-class constraints} for a given $n$ by 
\be 
\TopCon(n)                                     \m ,
\ee 
The {\it absence of constraints} is just $id$.  
The {\it space of classical first-class linear constraints} is 
\be 
\FlinCon(n)                                     \m ,
\ee 
and the {\it space of gauge constraints} is 
\be 
\GaugeCon(n)                                    \m . 
\ee 
\n Whereas $\bFlin = \bGauge$ in more commonly encountered cases, e.g.\ \cite{HTBook} contains a counterexample to these always being the same.  

\m 

\n{\bf Structure 2} The totality of constraints subalgebraic structures for a given formalism of a given theory  for each $n$, 
\be
\mbox{bounded lattice } \m \lattice_{\sCon(n)}      \m . 
\ee
The identity algebraic structure                            is the bottom alias zero element, whereas  
the top     algebraic structure of first-class constraints is the top    alias unit element. 
All other elements are middle elements: the {\it Z constraint algebraic structures}, denoted by 
\be 
\bFrZ(n) \m \mbox{ with each type indexed by } \m \fZ  \m . 
\ee
\n{\bf Remark 1} These are comparable to configuration spaces and phase spaces in the study of the nature of Physical Law, 
and whose detailed structure is needed to understand any given theory.
This refers in particular to the topological, differential and higher-level geometric structures observables algebraic structures support, 
now with also function space and algebraic levels of structure relevant. 

\m 

\n{\bf Remark 2} This means we need to pay attention to the Tensor Calculus on constraint algebraic structures as well, 
justifying our use of underbrackets to keep constraint-vectors distinct from spatial ones.  
%
%
%
{            \begin{figure}[!ht]
\centering
\includegraphics[width=1.0\textwidth]{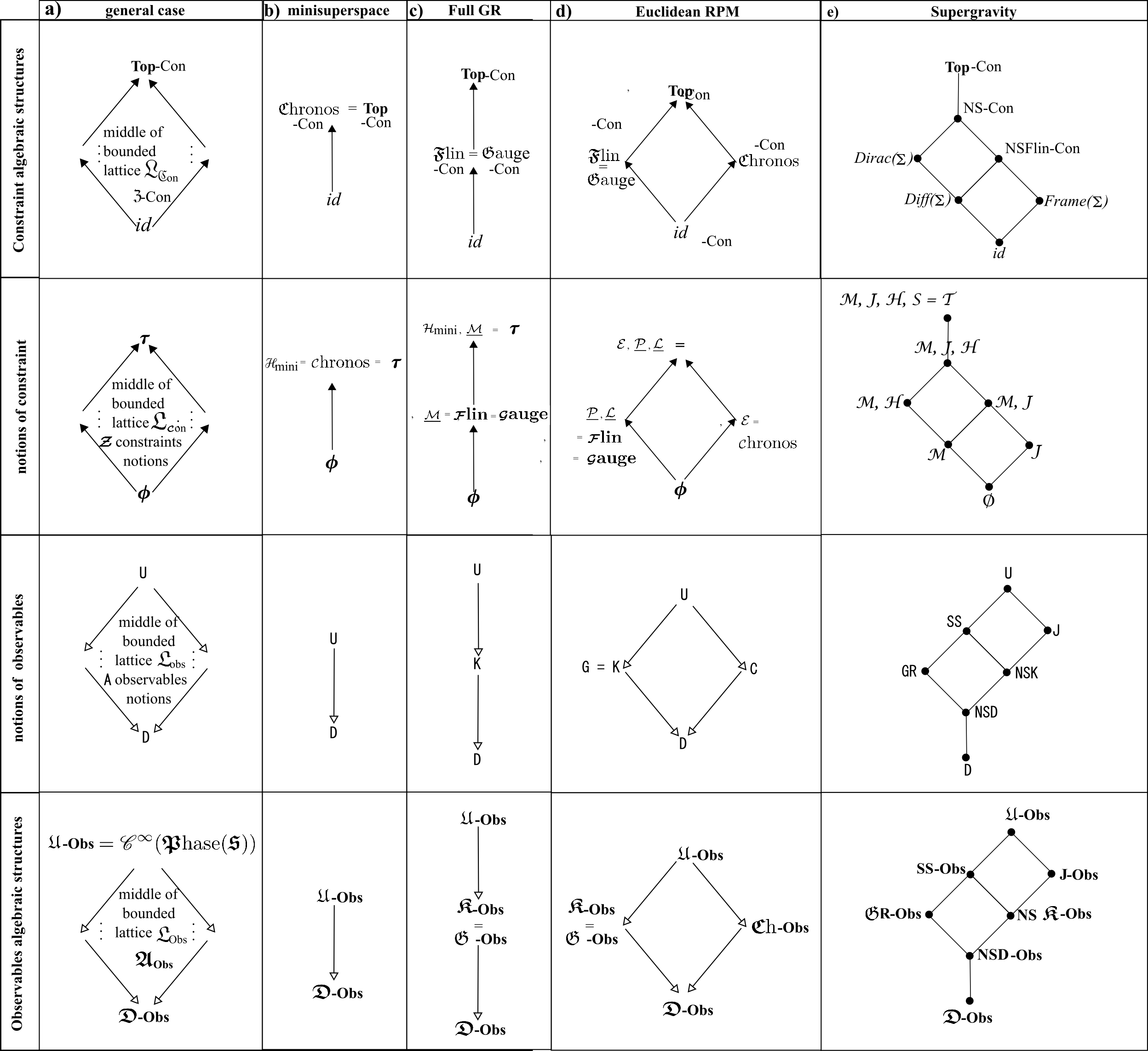}
\caption[Text der im Bilderverzeichnis auftaucht]{ \footnotesize{Lattices of notions                 of constraints (row 2)
                                                                      and of constraint subalgebraic structures     (row 1), 
												   with the dual lattices of notions                 of observables (row 3) 
                                                                      and of observables subalgebraic structures    (row 4).  
%
																	  
\m 
																	  
\n a) in general, schematically.  

\m 

\n b) For minisuperspace, i.e.\ spatially-homogenoeous, GR.

\m 

\n c) for full GR: a first arena with a nontrivial middle. 

\m 

\n d) for Euclidean relational particle mechanics \cite{FileR, ABook}, which serves as a first model arena with a nontrivial-poset middle rather than just a chain.  

\m 

\n e) Supergravity possesses a {\it non-supersymmetric first-class linear constraint algebra} $\NSFlin = \{ \bscJ , \, \bscM \}$, 
                             a {\it non-supersymmetric total constraint algebroid}, $\NSC = \{\bscJ, \bscM, \scH\}$, 
			                   {\it Locally Lorentz observables} $\bttJ$, 
                               {\it non-supersymmetric \K observables} $\bNsK$, and 
                               {\it non-supersymmetric Dirac observables} $\bNsD$.        } }
\label{C-Latt}\end{figure}            }

\m 

\n{\bf Remark 1} The third bracket in the Dirac algebroid of GR's constraints is, 
firstly, a demonstration that $\u{\bscM}$ is an integrability of $\scH$ \cite{MT72}.   

\m 

\n{\bf Example 2} Supergravity requires a first-order formulation -- such as beins $\bee$ in place of metrics $\bh$; 
$\bh$ is the product of two $\bee$, each of which carries one of the metric's indices and a flat spacetime index that is contracted over.  
This in turn brings in further `local Lorentz frame' linear constraints $\bscJ$.
Canonical supergravity requires furthermore use of graded Poisson brackets $\lp \m \cb \rp_{\sg}$. 
\be 
\lp  \scH \, \cb  \scH  \rp_{\sg}  \msim  \u{\bscM}  \times  \mbox{(structure functions)}  \mmp  ...
\label{H->M}
\ee 
still applies; first-order formulations of both GR and Supergravity furthermore have $\bscJ$ is an integrability of $\bscM$ 

\m 

\n The big result however is that the further linear supersymmetric constraint $\bscS$ has the quadratic Hamiltonian constraint $\scH$ as an integrability \cite{T77}, 
\be 
\lp  \scS  \cb  \scS  \rp  \msim  \scH  \mmp  ... 
\label{S->H}
\ee
Consequently, the $\flin$ do not close in this case. 

\m 

\n See Fig \ref{C-Latt}.e)  for some geometrically and physically significant subalgebraic structures that Supergravity possesses instead. 
by Supergravity.

\m 

\n (\ref{H->M}) and (\ref{S->H}) confirm that Supergravity's constraint algebraic structure is a superalgebroid.

\section{The Nambu Algorithm for Spacetime Generator Closure}\label{NASGC}

\n This parallels the Lie Algorithm: given an initial set of generators, do these close? 
This is now to be assessed with Nambu brackets rather than Nambu--Posson brackets, and has one case less than the next section's parallel of the Dirac Algorithm.  

\m 

\n{\bf Definition 1} We now extend Dirac's \cite{Dirac} notion of weak equality to mean equality up to additive linear functionals of the generators, 
and the portmanteau inequality likewise.

\subsection{Nambu's Little Algorithm}\label{NLA}

\n{\bf Definition 1} {\it `Nambu's Little Algorithm'}\footnote{Again, this is named in honour of Nambu, rather that being by Nambu.
It is a direct parallel of Lie's Algorithm \cite{Lie} as extended by \cite{Higher-Lie, XIV} Dirac's insights \cite{Dirac} but to the exclusion of appending}.  
%
This consists of evaluating Nambu brackets between a given input set of generators so as to determine whether these are consistent and complete.  
Three possible types of outcome are allowed in this setting.    

\m 

\n{\bf Type 0)} {\it Inconsistencies}.

\m 

\n{\bf Type 1)} {\it Mere identities}. 

\m 

\n{\bf Type 2)} {\it Further generators}.

\m 

\n{\bf Remark 1} If type 2) occurs (in the absence of type 0)'s immediate inconsistency), 
the resultant generators are fed into the next iteration of the Nambu Algorithm, which starts with the extended set of generators.  
One is to proceed thus recursively until one of the following termination conditions is attained, closely paralleling Sec \ref{NDLA}'s definitions.

\m  

\n{\bf Termination Condition 0} {\it Immediate inconsistency}.

\m 

\n{\bf Termination Condition 1} {\it Combinatorially critical cascade}. 

\m 

\n{\bf Termination Condition 2} {\it Sufficient cascade}. 

\m 

\n{\bf Termination Condition 3} {\it Completion}. 

\m 

\n{\bf Remark 2} Termination Conditions 0) and 2) are a matter of rejection of an initial candidate set of generators.  
The possibility of either of these applying at some subsequent iteration justifies our opening conception in terms of ab initio consistency.

\subsection{Full Nambu Algorithm}

\n First and second classness generalizes to candidate generators in the general Lie Theory \cite{XIV}, and then similarly also to the general Nambu Theory. 

\m 

\n Second-class generator removal also generalizes from the Dirac to the Lie setting \cite{XIV} and thus to the Nambu one too, 
provided that it is even-$n$ Nambu so as to permit inversion.  
 
\m 

\n In this algorithm, except that whenever Nambu-second-class generators appear, one switches to (new) Nambu--Dirac brackets that factor these in.  
This amounts to a fourth type of equation being possible, as follows.  

\m 

\n{\bf Type 4)} {\it Further Nambu-second-classness} can arise.

\m 

\n{\bf Type 5} {\it Discovery of a topological obstruction} to having a Nambu algebraic structure of generators.
The most obvious examples of this are anomalies \cite{Bertlmann} at the quantum level in the binary case; 
it is however a general brackets phenomenon rather than specifically a quantum phenomenon, and admits an $n$-ary Nambu counterpart

\subsection{Generator algebraic structures introduced}\label{NoG}

\n{\bf Structure 1} The end product of a successful candidate theory's passage through the Nambu Algorithm 
is a {\it Nambu algebraic structures of generators} consisting solely of first-class generators closing under Nambu (or more generally Nambu--Dirac) brackets. 

\m 

\n{\bf Modelling assumption} Assume that neither the topological nor tertiary complications explained below occur.

\m 

\n{\bf Structure 2} Under this condition, schematically, 
\beq
\ll  \bscg  \cb  \bscg  \rl  \peqs  0                                                                      \m .
\label{G-G}
\eeq
This is a portmanteau for the strong version
\beq
\ll  \bscg  \cb  \bscg  \rl   \es  0                                                                       \m , 
\label{G-G-S}
\eeq
and the weak version: 
\beq
\ll  \uc{\bscg}  \cb  \uc{\bscg}  \rl  \es  \uc{\uc{\uc{\biG}}} \cdot \uc{\bscg}                            \m .
\label{G-G-W}
\eeq
The $\biG$ here can be the structure constants of a Nambu algebra, or a Nambu algebroid's spacetime functions,  
\be 
\biG(\biS) \m . 
\ee 
\beq
\ll  \uc{\bscg}  \cb  \uc{\bscg}  \rl  \es  \uc{\uc{\uc{\biG}}} \cdot \uc{\bscg}  \mmp  
                                            \uc{\uc{\uc{\biH}}} \cdot \uc{\bsch}  \mmp 
								            \uc{\uc{\theta}}                       \m .
\label{F-F-W-4}
\eeq
as a more general output, with discovered generators. 
One then needs to look at $\ll  \bscg  \cb  \bsch  \rl$ 
                      and $\ll  \bsch  \cb  \bsch  \rl$.    

\m 

\n In the $n$-ary case, 
\beq
\ll  \uc{\bscg}  \cb ... \,  \cb  \uc{\bscg}  \rl  \es  \s{\uc{\biC}}{\uc{\mbox{\bf...}}}   \cdot \uc{\bscg}  \mmp 
                                                        \s{\uc{\biD}}{\uc{\mbox{\bf...}}}   \cdot \uc{\bsch}  \mmp
											            \s{\uc{\btheta}}{\uc{\mbox{\bf..}}}                   \m .
\label{F-F-W-5}
\eeq
where the double dot indicates that $\btheta$ is an $n$-array. 

\m 

\n{\bf Example 1} Spacetime also possesses its own version of closure -- now not Constraint Closure but Generator Closure, in particular for 
\be 
\lFrg_{\sS} = Diff(\Frm)
\ee 
-- spacetime diffeomorphisms -- the generators obey  
\be
\ls ( \, \vec{\bcalD} \, | \, \vec{\bX}) \, \cb ( \, \vec{\bcalD} \, | \, \vec{\bY} \, )  \rs  \es  
    ( \, \vec{\bcalD} \, | \, \s{\longrightarrow}{ [ \, \bX, \, \bY \, ]  } \, )              \m .  
\label{Sp-Diff}
\ee
$\vec{\bX}$ and $\vec{\bY}$ are here spacetime smearing variables, whereas $[ \m , \, \m ]$ is Differential Geometry's commutator of two vectors.   
Our equation is a subcase of generator-weakly vanishing. 

\m 

\n{\bf Example 2} Supergravity's spacetime super-diffeomorphisms close schematically as  
\be
\ls  ( \,  \vec{\bcalS\bcalD} \, | \, \vec{\bX} ) \, \cb ( \vec{\bcalS\bcalD} \, | \, \vec{\bY} \, )  \rs_{\sg}  \es  
     ( \,  \vec{\bcalS\bcalD} \, | \, \s{\longrightarrow}{ [ \, \bX , \,  \bY \, ]}_{\sg} \, )
\ee
for subscript $\mg$ denoting grading.

\subsection{Lattices of Nambu subalgebraic structures of generators}

\n{\bf Structure 1} There is further interest in finding the (in particular conceptually meaningful) 
consistent subalgebras supported by the top generators algebra. 
The notions of generator algebraic structure in question form a bounded lattice for each $n$, 
\be 
\lattice_{\sgen(n)}
\ee 
with a complete consistent set of generators $\gen(n)$ as top element, and the absence of generators $\emptyset$ as bottom element.
The other members of the lattice of notions of generator algebraic structure are middle elements, the {\it notions of Z-algebraic structure of generators} 
denoted by 
\be 
\bscz(n) \m \mbox{ with each type indexed by } \m  \fz  \m .
\ee  
See row 2 of Fig \ref{C-Latt} for a schematic sketch. 

\m

\n{\bf Structure 1}  The {\it top space of first-class linear generators} is  
\be 
\top(n)                \m ,
\ee 
The {\it space of absence of generators} is just $id$.  

\m 

\n The {\it space of classical first-class linear generators} is 
\be 
\FlinGen(n)            \m ,
\ee 
and the {\it space of gauge generators} is 
\be 
\GaugeGen(n)           \m . 
\ee 
For $n = 2$, we use no number at all due to coinciding with the usual Lie case.

\m 

\n {\bf Structure 2} The totality of generator subalgebraic structures for a given formalism of a given theory with a given $n$,  
\be
\mbox{bounded lattice } \m \lattice_{\sGen(n)}   \m . 
\ee
The identity algebraic structure is the bottom alias zero element, and 
the top algebraic structure of first-class generators is the top alias unit element. 
All other elements are middle elements: the {\it Z generator algebraic structures}, denoted by 
\be 
\bFrZ(n) \m \mbox{ with each type indexed by } \m \fz  \m . 
\ee

\section{Nambu constrained observables}\label{NO}

\subsection{Unrestricted observables}\label{U-Can}

{\bf Structure 1} The most primary notion of observables involves {\it Taking a Function Space Thereover}.  
In the classical binary Poisson brackets setting, this refers to over $\Phase(\lFrs)$.  
The first such output are {\it unrestricted observables}, 
\be 
\sbiU(\biQ, \biP)  \m . 
\ee 

\n{\bf Structure 2} These form some function space of suitably-smooth functions, such as 
\be 
\UnresObs  =  \FrC^{\infty}(\Phase(\lFrs))  \m .
\ee 
At the classical level, this is a trivial extension of the theoretical framework; 
at the quantum level, however, self-adjointness and Kinematical Quantization already impinge at this stage.

\m   

\n{\bf Structure 3} For $n$-ary Nambu Theory, {\it unrestricted observables} are  
\be 
\sbiU(\biQ, \u{\, \biP\,})  \m ,  
\ee 
forming the space 
\be 
\UnresObs(n)  \es  \FrC^{\infty}(\Phase(\lFrs, n)))  \m . 
\ee 
While no brackets relations are imposed, these already succeed in being different from the binary Poisson case's because of now being functions over  
the distinctive $n$-ary generalized phase space.

\subsection{Constrained observables}\label{CA}

\n We next consider {\bf generalized phase space functions} restricted by {\bf zero-commutant Nambu brackets with the Nambu-first-class constraints} 

\m 

\n{\bf Motivation 1} In a constrained theory, 
constrained observables are more physically useful than just any functions (or functionals) of $\biQ$ and $\biP$, 
due to their containing between more, and solely, physical information.    

\m 

\n The $n = 2$ binary subcase (Dirac subcase of Lie) is covered in \cite{DiracObs, K93, ABook, VIII}.   

\m 

\n{\bf Structure 1} In the binary case, $\con$ imposes Lie bracket {\it commutant conditions} on observables, 
\be 
\lp  \con  \cb  \Ob  \rp  \peqs  0  
\ee
on the canonical observables functions, $\Ob$. 
I.e.\
\be 
\lp  \con  \cb  \Ob  \rp  \es   0  
\ee
in the strong case, or 
\be  
\lp  \uc{\con}  \cb  \uo{\Ob}  \rp  \es  \uc{\uo{\uc{\biW}}} \cdot \uc{\sbiC}  
\ee 
in the weak case, with structure constants $\biW$. 

\m 

\n{\bf Observables Theorem} \cite{AObs} 

\m

\n i) For this commutant condition to be consistent \cite{AObs}, 
the $\con$ must moreover constitute a closed subalgebraic structure of first-class constraints $\bscF$.  

\m 

\n ii) These binary notions of observables themselves close to form further Lie algebras.  

\m 

\n{\u{Proof}} i) This follows from the `OCC' Jacobi identity 
\be
  \lp  \Ob   \cb  \lp \con  \cb  \con  \rp  \rp  \es 
- \lp  \con  \cb  \lp \con  \cb  \Ob   \rp  \rp 
- \lp  \con  \cb  \lp \Ob   \cb  \con  \rp  \rp  \peqs  0     \m .  
\ee  
ii) This follows from the `COO' Jacobi identity
\be
  \lp  \con  \cb  \lp  \Ob   \cb  \Ob   \rp  \rp   \es 
- \lp  \Ob   \cb  \lp  \Ob   \cb  \con  \rp  \rp 
- \lp  \Ob   \cb  \lp  \con  \cb  \Ob   \rp  \rp  \peqs  0     \m . \m \Box
\ee
\n{\bf Remark 1} i) {\sl decouples} Assignment of Canonical Observables to occur {\sl after} establishing Constraint Closure of the constraints.  

\m 

\n{\bf Structure 2} In the ternary case, four kinds of observables are a priori possible \cite{AObs}:
\be 
\lp  \con  \cb  \con   \cb  \Ob_1   \rp  \peqs  0  \m , 
\ee
\be 
\lp  \con  \cb  \Ob_2  \cb  \Ob_2   \rp  \peqs  0  \m , 
\ee
\be 
\lp  \con  \cb  \Ob_1  \cb  \BOb_1  \rp  \peqs  0  \m ,    
\ee
\be 
\lp  \con  \cb  \Ob_2  \cb  \BOb_2  \rp  \peqs  0  \m ,    
\ee
$\BOb$ denotes `biobservables': Nambu-commuting with one $\sbiC$ and one $\Ob$, compare `binormal' in the context of basic geometry of curves in 3-$d$. 
Let us denote the spaces each if these form by $\Obs_1$, $\Obs_2$, $\BiObs_1$, $\BiObs_2$ respectively.  

\vspace{10in}

\n{\bf Nambu Observables Non-Proliferation Theorem} 

\m 

\n i)   All kinds of observables in the arbitrary-$n$ extension of the above sense in fact coincide. 

\m 

\n ii)  The resulting unique kind of observable supported by the $n$-ary Nambu bracket necessitates a closed Nambu algebraic structure of constraints.  

\m 

\n iii) This coincident kind itself closes as a Nambu algebraic structure.  

\m 

\n{\u{Proof}} Repeatedly use the fundamental identity, as follows.

\m  

\n In the ternary case,  
\be 
\lp  \con  \cb  \con  \cb  \lp  \con  \cb  \con  \cb  \Ob_1  \rp         \rp             \es  
\lp        \lp  \con  \cb       \con  \cb  \con  \rp  \con   \cb  \Ob_1  \rp             \mmp 
\lp  \con  \cb        \lp       \con  \cb  \con  \cb  \con   \rp         \cb  \Ob_1 \rp  \mmp
\lp  \con  \cb  \con  \lp       \con  \cb  \con  \cb  \Ob_1  \rp         \rp             \m .
\ee
\be 
\Rightarrow \m  \lp  \con  \cb  \con  \cb  \con  \rp  \peq  0  \m , 
\ee
so the $\Ob_1$ necessitate a closed Nambu subalgebraic structure of constraints. 
\be 
\lp  \con  \cb  \Ob_1  \cb         \lp   \con   \cb  \con   \cb  \Ob_1  \rp         \rp  \es  
\lp        \lp  \con   \cb  \Ob_1  \cb   \con   \rp  \con   \cb  \Ob_1  \rp              \mmp 
\lp  \con  \cb         \lp  \con   \cb   \Ob_1  \cb  \con   \rp         \cb  \Ob_1  \rp  \mmp
\lp  \con  \cb  \con   \lp  \con   \cb   \Ob_1  \cb  \Ob_1  \rp         \rp              \m .
\ee
\be 
\Rightarrow \m  \lp  \con  \cb  \Ob_1  \cb  \Ob_1  \rp  \peq  0  \m , 
\ee 
\be 
\Rightarrow \m  \Obs_1  \msubseteq  \Obs_2  \m . 
\label{f->s}
\ee 
\be 
\lp  \Ob_1 \cb  \Ob_1  \cb         \lp  \con   \cb  \con   \cb  \Ob_1  \rp  \rp        \es  
\lp        \lp  \Ob_1  \cb  \Ob_1  \cb  \con   \rp  \con   \cb  \Ob_1  \rp             \mmp 
\lp  \con  \cb         \lp  \Ob_1  \cb  \Ob_1  \cb  \con   \rp         \cb  \Ob_1 \rp  \mmp
\lp  \con  \cb  \con   \lp  \Ob_1  \cb  \Ob_1  \cb  \Ob_1  \rp         \rp             \m .
\ee
\be 
\Rightarrow \m  \lp  \Ob_1  \cb  \Ob_1  \cb  \Ob_1  \rp   \peq  0  \m : 
\ee 
so the $\Ob_1$ form a closed Nambu subalgebraic structure of observables.   

\m 

\n For $\Ob_2$, 
\be 
\lp  \con   \cb  \Ob_2  \cb         \lp  \Ob_2  \cb  \Ob_2  \cb  \con  \rp       \rp  \es  
\lp         \lp  \con   \cb  \Ob_2  \cb  \Ob_2  \rp  \Ob_2  \cb  \con  \rp            \mmp 
\lp  \Ob_2  \cb         \lp  \con   \cb  \Ob_2  \cb  \Ob_2  \rp        \cb  \con \rp  \mmp
\lp  \Ob_2  \cb  \Ob_2  \lp  \con   \cb  \Ob_2  \cb  \con   \rp        \rp            \m .
\ee
\be 
\Rightarrow \m  \lp  \con  \cb  \Ob_2  \cb  \con  \rp  \peq  0    \m : 
\ee
\be 
\Rightarrow \m  \Obs_1  \msupseteq  \Obs_2  \m . 
\label{s->f}
\ee 
Together, (\ref{f->s}, \ref{s->f}) $\Rightarrow$
\be 
\Obs_1  \es  \Obs_2  \m , 
\ee  
by which also 
\be 
\BiObs_1  \es  \BiObs_2  \m .  
\ee 
\n For the third kind, its definition already entails $\Ob$ being an observable of the joint first and second kind, and the underlying $\con$'s closing. 
\be 
\lp  \con  \cb  \con  \cb        \lp  \con  \cb  \Ob   \cb  \BOb  \rp        \rp  \es  
\lp        \lp  \con  \cb  \con  \cb  \con  \rp  \Ob   \cb  \BOb  \rp             \mmp 
\lp  \con  \cb        \lp  \con  \cb  \con  \cb  \Ob   \rp        \cb  \BOb  \rp  \mmp
\lp  \con  \cb  \Ob   \cb        \lp  \con  \cb  \con  \cb  \BOb  \rp        \rp  \m .
\label{1-of-6}
\ee
\be 
\Rightarrow \m  \lp  \con  \cb  \con  \cb  \BOb  \rp  \peqs  0 
\ee   
\be 
\Rightarrow \m  \BiObs  \msubseteq  \Obs              \m . 
\label{t->f}
\ee 
But any $\Ob$ obeys 
\be 
\lp  \con  \cb  \Ob  \cb  \Ob  \rp  \es  0  \mma 
\ee 
so any $\Ob$ serves as a $\BOb$. 
Thus 
\be 
\BiObs  \msupseteq  \Obs  \m . 
\label{f->t}
\ee 
Together, (\ref{t->f}, \ref{f->s}) give that 
\be 
\BiObs    \es     \Obs  \m . 
\ee
%

\m

\n The general $n$ case firstly uses all the fundamental identities for the leading case. 
Then it composes each other kind of observables-defining bracket with all $\con$'s to form further fundamental identities 
to complete a `two way inclusion proof' in parallel to the above layout.   $\Box$ 

\m 

\n{\bf Remark 2} In the weak ternary case, we have explicity 
\be 
\lp  \uc{\con}  \cb  \uc{\con}  \cb  \uo{\Ob}  \rp  \es  \uc{\uo{\uc{\uc{\biW}}}} \, \uc{\sbiC}   \m , 
\ee
for 4-array $\biW$.

\subsection{Dirac and Nambu--Dirac observables as top observables}

\n{\bf Structure 1} The opposite extreme to imposing no restrictions is to impose all of a modelling situation's first-class generators.
This returns the {\it top observables} $\Top(n)$. 
In the binary case, these obey 
\beq
\lp  \bscF  \cb  \Top  \rp  \peqs   0  \m , 
\label{C-D}
\eeq 
and are alias {\it Dirac observables} \cite{DiracObs}.
\n In the general $n$-ary case, these obey 
\beq
\lp  \bscF(n)  \cb  ... \,  \bscF(n)  \cb  \Top(n)  \rp  \peq  0  \m .
\label{C-D(n)}
\eeq
and could also be termed {\it Nambu--Dirac observables}.

\m 

\n{\bf Remark 1} The unrestricted and (Nambu--)Dirac notions of observables are universal over all physical theories.

\subsection{Middling notions of canonical observables}

\n{\bf Structure 1} Some physical theories moreover support further middling notions of classical canonical observables. 
These correspond to each closed subalgebraic structure the constraint algebraic structure possesses. 

\m  

\n{\bf Definition 1} {\it First-class linear observables} alias {\it \K observables} $\Kuchar$ \cite{K93} are functions over phase space for which  
\beq
\lp  \bFlin  \cb  \Kuchar  \rp  \peqs  0    \m .
\label{FLIN-K}
\eeq
\n{\bf Definition 2} $\lFrg$-observables alias {\it gauge-invariant quantities} $\gauge$ are functions over phase space for which  
\beq
\lp \bGauge  \cb  \gauge  \rp  \peqs  0     \m .
\label{GAUGE-G2}
\eeq
\n{\bf Definition 3} In cases in which quadratic constraints $\Chronos$, such as the energy constraint of Mechanics or the Hamiltonian constraint of GR,
 a notion of {\it Chronos observables} may become meaningful, i.e.\ functions over phase space obeying 
\beq
\lp  \Chronos  \cb  \chronos  \rp  \peqs  0      \m .
\label{CHRONOS-C}
\eeq
{\bf Example 1} GR supports \K but not Chronos observables, since $\u{\bscM}$ closes but $\scH$ does not, having $\u{\bscM}$ as an integrability (see Fig \ref{C-Latt}.c).   

\m

{\bf Example 2} Relational Particle Mechanics \cite{FileR, ABook}, however, supports both \K and Chronos observables (see Fig \ref{C-Latt}.d).  

\m 

\n{\bf Remark 1} Kucha\v{r}'s notion of observables does not however have theory-independent significance. 
This role is, rather, replaced by whatever non-extremal elements the bounded lattice of notions of observables of a theory happens to possess, 
i.e.\ Sec \ref{CA}'s notion of A-observables.    

\m  

\n{\bf Example 3} Consult Fig \ref{C-Latt}.e) for further middling notions of observables supported by Supergravity.   

\m 

\n{\bf Definition 4} {\it First-class Nambu observables}, which I also term {\it Nambu--\K observables} $\Kuchar(n)$ are functions over generalized phase space 
for which 
\beq
\lp  \bFlin(n)  \cb  ... \,  \bFlin(n)  \cb  \Kuchar(n)  \rp  \peqs  0  \m .
\label{FLIN-K2}
\eeq
\n{\bf Definition 5} $\lFrg$-{\it Nambu observables} alias {\it gauge-invariant quantities} $\gauge(n)$ are functions over generalized phase space for which 
\beq
\lp  \bGauge(n)  \cb  ... \,  \bGauge(n)  \cb  \gauge(n)  \rp  \peqs  0   \m .
\label{GAUGE-G}
\eeq
\n{\bf Definition 6} In cases in which $\Chronos(n)$ self-Nambu-closes, a notion of {\it Nambu--Chronos} observables $\chronos(n)$ becomes meaningful, obeying 
\beq
\lp  \Chronos(n) \cb  ... \, \Chronos(n)  \rp  \chronos(n)  \rp  \peqs  0  \m .
\label{CHRONOS-C2}
\eeq
\n {\bf Structure 2} For each $n$, the totality of notions of canonical observables form a 
\be
\mbox{bounded lattice } \m \LattObs(n) \m \mbox{ dual to that of constraint algebras} \mma  \LattCon(n)  \m .
\ee  
Its top and bottom elements are $\Top(n)$ and $\Unres(n)$ notions of observables, 
whereas the middle elements are {\it Nambu--A observables}, $\bttA(n)$ with each type of such a theory supports indexed by $\fZ$.

\subsection{Observables algebraic structures}\label{OAS}

\n{\bf Structure 1}  We denote the {\it space of top alias Nambu--Dirac observables} is 
\be 
\TopObs(\lFrs, n)   \m ,
\ee 
{\it space of first-class linear alias Nambu--Kucha\v{r} observables} is 
\be 
\FlinObs(\lFrs, n)   \m ,
\ee 
and {\it space of gauge} alias $\lFrg$-{\it Nambu observables} is 
\be 
\GaugeObs(\lFrs, n)  \m .
\ee 
{\bf Structure 2} For each $n$, the totality of notions of observables form a 
\be
\mbox{bounded lattice } \m \LattObs(n) \m \mbox{ dual to that of constraint algebraic structures} \mma  \LattCon(n)  \m .
\ee
The algebraic structure of unrestricted observables is the top alias unit element, and 
the algebraic structure of Dirac        observables is the bottom alias zero element. 
All other elements are middle elements: the {\it Nambu--A observables algebraic structures}, denoted by 
\be 
\lFra(n) \m \mbox{ with each type indexed by} \m \fZ  \m . 
\ee
\n{\bf Remark 1} The above spaces, like the constraint algebraic structures, 
are comparable to configuration spaces and phase spaces in the study of the nature of Physical Law, 
and whose detailed structure is needed to understand any given theory.
This refers in particular to the topological, differential and higher-level geometric structures observables algebraic structures support, 
now with also function space and algebraic levels of structure relevant. 

\m 

\n{\bf Remark 2} This means we need to pay attention to the Tensor Calculus on observables algebraic structures as well, 
justifying our use of undertildes to keep observables-vectors distinct from constraints ones and spatial ones.  
In fact, the increased abstraction of each of these is `indexed' by my notation: no turns for spatial, one for constraints and two for observables.   

\m 

\n{\bf Remark 3} The sizes of the spaces run in the dual lattice pair run in opposition. 
I.e.\ the bigger a constraint algebraic structure, the smaller the corresponding space of observables is. 
This is clear enough from constraints acting as restrictions, adding PDEs that the observables must satisfy...

\subsection{Explicit solution for constrained observables} 

\n This brackets equation can moreover be recast as explicit PDEs to be solved using the Flow Method \cite{Lie, Gug63, John, Olver2, Lee2, Olver}. 
This particular application essentially constitutes a generalization of {\it Lie's Integral Approach to Geometrical Invariants} \cite{Lie, Gug63, DO-1}, 
elevated successively to phase space, and then to generalized phase space as equipped with Nambu brackets.
And finally uplifted moreover to a free characteristic problem for finding 
'suitably-smooth functions of phase space invariants', i.e.\ observables.  

\m

\n{\bf Structure 1} More specifically, observables' defining relations can be recast \cite{ABook, DO-1} 
                    for practical purposes of solution as flow PDES \cite{John, Lee2}.  

\m 
					
\n A first-order quasilinear PDE 
\be 
\sum_{A} a^{A}(x^{B}, \phi) \pa_A \phi  \es  b(x^{B}, \phi)
\ee 
for unknown variable $\phi$ and individually-arbitrary indices $A$ and $B$ is thus recast as a flow system of ODEs 
\be 
\dot{x}^{A}  =  a^{A}(x^{B}, \phi)  \mma  
\ee 
\be 
\dot{\phi}   =      b(x^{B}, \phi)  \m .  
\ee
Here $\dot := \pa/\pa\tau$ for some parameter $\tau$ along the flow.  

\m 

\n We more generally entertain a system of $A = 1$ to $M$ such PDEs, 
\be 
\sum_{B = 1}^N a^{AB}(x^{C}, \phi) \pa_B \phi = b^A(x^{C}, \phi)
\ee
There is no general method for such, the outcome depending firstly on determinedness, and secondly on integrability.  

\m 

\n{\bf Remark 1} In e.g.\ the ternary case (see \cite{DO-1} for binary case), 
we arrive at our observables PDE system by the Nambu bracket being an alternating sum of six terms of the form 
\be 
\frac{ \pa \con }{ \pa \biQ }  \frac{ \pa \con }{ \pa \u{\,\biP\,} }  \frac{ \pa \Ob }{ \pa \u{\,\biP\,}^{\prime} }  
\ee 
or some permutation of the denominators. 
I.e.\, Summing over each multi-index of $\biQ$,
$$
0  \peq  \{ \con , \, \con^{\prime} , \, \Ob \}  
    \es  \left|  \frac{ \pa (\con , \, \con^{\prime} , \, \Ob) }{ \pa (\biQ, \, \biP_1, \, \biP_2) }  \right|
    \es  \left|  \frac{ \pa (\con , \, \con^{\prime} , \, -  ) }{ \pa (\biQ, \, \biP_1, \, \biP_2) }  \right| \, \Ob \m ,  
$$
where the last form makes explicit linearity of the PDE system (over all independent values of the constraints' two indices) and  -- is a blank slot.    
But the $\con$ are known expressions, so this leaves us with a first-order differential equation in $\Ob$.

\n{\bf Structure 2} Concretely, on the one hand the strong case of this is the homogeneous-linear PDE system 
\be
\left(    \frac{\pa \con}{\pa \biQ}   \left(        \frac{ \pa \con^{\prime} }{ \pa \biP_1 } \frac{ \pa  }{ \pa \biP_2 } 
                                              \mmm  \frac{ \pa \con^{\prime} }{ \pa \biP_2 } \frac{ \pa  }{ \pa \biP_1 }  \right)  \mmm 
          \frac{\pa \con}{\pa \biP_1} \left(        \frac{ \pa \con^{\prime} }{ \pa \biP_2 } \frac{ \pa  }{ \pa \biQ} 
                                              \mmm  \frac{ \pa \con^{\prime} }{ \pa \biQ }   \frac{ \pa  }{ \pa \biP_2 }  \right)  \mmp 
          \frac{\pa \con}{\pa \biP_2} \left(        \frac{ \pa \con^{\prime} }{ \pa \biQ }   \frac{ \pa  }{ \pa \biP_1 } 
                                              \mmm  \frac{ \pa \con^{\prime} }{ \pa \biP_1 } \frac{ \pa  }{ \pa \biQ }    \right)		  \right)  \Ob  \es  0  \m .						
\ee
On the other hand, the weak case of this is the inhomogeneous-linear PDE system
\be
\left(    \frac{\pa \con}{\pa \biQ}   \left(        \frac{ \pa \con^{\prime} }{ \pa \biP_1 } \frac{ \pa }{ \pa \biP_2 } 
                                              \mmm  \frac{ \pa \con^{\prime} }{ \pa \biP_2 } \frac{ \pa }{ \pa \biP_1 }  \right) \mmm 
          \frac{\pa \con}{\pa \biP_1} \left(        \frac{ \pa \con^{\prime} }{ \pa \biP_2 } \frac{ \pa }{ \pa \biQ } 
                                              \mmm  \frac{ \pa \con^{\prime} }{ \pa \biQ }   \frac{ \pa }{ \pa \biP_2 }  \right) \mmp 
          \frac{\pa \con}{\pa \biP_2} \left(        \frac{ \pa \con^{\prime} }{ \pa \biQ }   \frac{ \pa }{ \pa \biP_1 } 
                                              \mmm  \frac{ \pa \con^{\prime} }{ \pa \biP_1 } \frac{ \pa }{ \pa \biQ }    \right)		   \right) \Ob  \es
\s{\uc{W}}{\uc{\uo{. \m .}}} \, \uc{\con}											                                                                    \m .						
\ee
\n{\bf Remark 2} \cite{DO-1, VIII}'s argument in the binary case for a free characteristic problem treatment of observables PDE systems 
transcends to our Nambu setting.  

\m 

\n{\bf Remark 3} So does the binary case's guarantee of integrability, by which solution is always in principle possible. 

\m 

\n{\bf Remark 4} So does finally the approach by the Flow Method sequentially: equation by equation.

\m 
				 
\n{\bf Remark 5} Demanding entire algebras rather than individual solutions to such moreover of course involves further Nambu brackets algebra level checks.  
Moving up the lattice moreover amounts to successive restrictions of the unreduced problem's 'solution manifold' function space, 
corresponding to applying larger and larger consistent subsets of constraints.

\section{Nambu Spacetime observables}\label{NSO}

\subsection{Unrestricted spacetime observables}\label{U-Sp}

The most primary notion of observables involves {\it Taking a Function Space Thereover}.  

\m 

\n In the classical spacetime and unrestricted setting, this refers to over $\PRiem(\FrM)$: the space of spacetimes on a fixed spacetime topology $\FrM$.    

\m 

\n{\bf Structure 1} Let us denote {\it unrestricted spacetime observables} by 
\be 
\uob(\biS)  \m . 
\ee 
for `spacetime configuration' functions $\biS$.
Nambu such are no different than Lie such, since no brackets conditions are imposed, nor is the underlying base space altered.  

\m 

\n{\bf Structure 2} These form some function space of suitably-smooth functions, such as 
\be 
\UnresObs(\lFrs) = \FrC^{\infty}(\PRiem(\FrS)))  \m .
\ee 
At the classical level, this is once again (c.f. Sec \ref{U-Can}) a trivial extension of the theoretical framework.

\subsection{Restricted spacetime observables}\label{RSO}

\n We next consider {\bf generalized phase space functions} restricted by {\bf zero-commutant Nambu brackets with the Nambu-first-class constraints} 

\m 

\n{\bf Motivation 1} In a theory with a group $\lFrg_{\sS}$ of physically irrelevant transformations on spacetime, 
restricted observables are more physically useful than just any functions (or functionals) of $\biS$, 
due to their containing somewhere between more, and solely, physical information.    

\m 

\n{\bf Structure 1} The $n = 2$ binary subcase (Dirac subcase of Lie) is covered in \cite{DiracObs, K93, ABook, VIII}.   
Here $\bscg$ imposes Lie bracket {\it commutant conditions} on observables, 
\be 
\ls  \bscg  \cb  \Ob  \rs  \peqs  0  \m .    
\ee
on the canonical observables functions, $\Ob$. 

\m 

\n{\bf Spacetime Observables Theorem} \cite{AObs} 

\m

\n i) For this commutant condition to be consistent \cite{AObs}, 
the $\con$ must moreover constitute a closed subalgebraic structure of first-class generators $\bscg$.  

\m 

\n ii) These binary notions of spacetime observables themselves close to form further Lie algebras.

\subsection{Restricted Nambu spacetime observables} 

{\bf Structure 1} We now consider entities which Nambu-commute with spacetime's generators: {\it Nambu spacetime observables}

\m 

\n{\bf Nambu Spacetime Observables Non-Proliferation Theorem} 

\m 

\n i)   There is only one kind (in Sec 3's sense) of Nambu spacetime observables per $n$-arity.

\m 

\n ii)  The resulting unique kind of spacetimes observables supported by the $n$-ary Nambu bracket necessitates a closed Nambu algebraic structure of generators.  

\m 

\n iii) This coincident kind itself closes as a Nambu algebraic structure.  

\m 

\n{\bf Example 1} In particular, one can place functions over spacetime which are $Diff(\FrM)$-invariant, 
i.e.\ Lie-brackets commutants with with $Diff(\FrM)$'s generators, 
\be
\ls( \, \bcalD \, | \, \bY  \, )  \cb ( \, \biS \, | \, \bZ \, ) \rs  \peqs   0 \m  
\label{Sp-Obs} 
\ee
for spacetime smearing variables $\bY$ and $\bZ$, or perhaps the generator-weak equality extension of this equation.  

\m 

\n{\bf Structure 2} These $\biS$ form an infinite-$d$ Nambu algebra 
\be 
\LFrS(n) = Diff(\Frm, n)  \m 
\ee 
(so the nature of diffeomorphisms changes in models resting on nontrivially-Nambu brackets).  

\m 

\n{\bf Remark 1} In the weak ternary case, we have explicity 
\be 
\ls  \uc{\bscg}  \cb  \uc{\bscg}  \cb  \uo{\ob}  \rs  \es  \uc{\uo{\uc{\uc{\biW}}}} \, \uc{\sbig}   \m , 
\ee
for 4-array $\biW$.

\subsection{Top spacetime observables} 

These use all of $\lFrg_{\sS}$. 
These are the analogue in this order-theoretic top sense to Dirac observables in the canonical case.
\n In the general $n$-ary case, this returns the {\it top spacetime observables} $\Dirac(n)$ obeying 
\beq
\ls  \bscf(n)    \cb  ... \,  \bscf(n) \cb  \Dirac(n)  \rs  \peqs   0   \m .
\label{C-D(n)-2}
\eeq
The unrestricted and top notions of spacetime observables are universal over all physical theories.

\subsection{Middling notions of spacetime observables}

\n{\bf Structure 1} Some physical theories moreover support further middling notions of spacetime observables. 
These correspond to each closed subalgebraic structure the spacetime algebraic structure possesses. 

\m  

\n{\bf Definition 1} {\it First-class linear spacetime observables} $\bttf(n)$ are quantities obeying  
\beq
\ls  \flin  \cb  ... \,  \cb  \flin  \cb  \bttf(n)  \rs  \peqs  0  \m .
\label{FLIN-K3}
\eeq
\n{\bf Definition 2} {\it Spacetime gauge observables} alias {\it spacetime gauge-invariant quantities} $\gaugeobs(n)$ are those for which 
\beq
\ls  \gauge  \cb  ... \,  \cb  \gauge  \cb  \gaugeobs(n)  \rs  \peqs  0  \m .
\label{GAUGE-G3}
\eeq
\n{\bf Example 1} {\it Diffeomorphism-invariant spacetime observables} are quantities forming zero Nambu spacetime brackets with all of a given theory's spacetime diffeomorphism generators, 
These are not necessarily top observables, or even the totality of the gauge observables, as exemplified by the Einstein--Maxwell Theory's 
top spacetime observables additionally forming zero brackets with the $U(1)$ gauge generators.  

\m 

\n {\bf Structure 2} For each $n$, the totality of notions of spacetime observables form a 
\be
\mbox{bounded lattice } \m \lattice_{\sobs(n)} \m \mbox{ dual to that of generator algebraic structures} \mma  \lattice_{\sgen(n)}  \m .
\ee  
Its top and bottom elements are     $\top(n)$ and $\unres$ respectively, 
whereas the middle elements are $\btta(n)$-{\it observables} with each type of such a theory supports indexed by $\fZ$.

\subsection{Solving PDEs from representing generators differentially}   

\n In this case, we obtain PDEs via representing generators differentially.
See \cite{PE-1} for the $n = 2$ Lie case of this (in a spatial geometry rather than spacetime geometry setting).

\subsection{Expression in Terms of Observables}

This subsection refers to whichever of the canonical or spacetime incarnations of observables.

\m 
			
\n{\bf Structure 1} Having an observables algebraic structure as a Function Space Thereover 
does not yet mean being able to express each physically-meaningful quantity in terms of observables.  
This involves the further step \cite{ABook, III} of eliminating out the ways each quantity is formulated in terms of unphysical quantities or mixed physical-unphysical quantities.
One can envisage this as an Algebra or Calculus based equation-solving process. 

\m 

\n{\bf Structure 2} For some physical quantities, some subset of observables may suffice rather than the whole space of these. 

\m 

\n{\bf Structure 3} It is additionally often convenient to pick a basis of Nambu observables in terms of which to express one's physical quantities. 
Consideration of functional independence, and bases, for observables algebraic structures is thus also a significant part of the theory of observables.

\section{Higher Nambu Theory}\label{NRR}

This parallels `higher Lie Theory' in the sense of \cite{Higher-Lie}.

\subsection{Lie Deformations and Lie Rigidity}\label{LDR} 

\n The `passing families of theories through the Dirac Algorithm' approach to Spacetime Construction from space 
\cite{RWR, AM13, ABook, A-Brackets, IX}, and to obtaining more structure from less for each of space and spacetime separately, 
are identified as \cite{Higher-Lie} Lie algebraic structure deformation procedures \cite{G64} that work by Lie Rigidity \cite{G64, NR66, CM} being encountered.

\m 

\n{\bf Structure 1} We denote {\it generator deformations} \cite{G64} by  
\be 
\uc{\bscg}  \mlongrightarrow  \uc{\bscg}\mbox{}_{\sbalpha}  \es  \uc{\bscg} \mmp  \uc{\uc{\balpha}} \cdot \uc{\bphi}  \m . 
\label{def}
\ee
$\balpha$ here in general carries a multi-index, so one has the corresponding multi-index inner product with an equally multi-indexed set of functions $\bphi$. 
These deformed generators can be viewed as terminating at linear order in $\balpha$   (rather than necessarily being small), 
                                              since each component of our $\balpha$'s is typically a priori real-valued.  

\m 

\n{\bf Remark 1} Nijenhuis and Richardson \cite{NR66} specialized Gerstenhaber's considerations of deformations \cite{G64} to the Lie algebraic setting. 
This work additionally attributes local stability under deformations to rigid Lie algebras. 

\m 

\n{\bf Structure 2} Gerstenhaber proceeds by placing a cohomological underpinning on rigidity results, 
which Nijenhuis and Richardson \cite{NR66} again specialize to the Lie algebra case as 
\be 
\mH^2(\Frg, \, \Frg) = 0  \m  \mbox{ diagnoses rigidity} \m .
\ee
\n{\bf Structure 3} This deformation approach has furthermore been generalized to ternary and $n$-ary Nambu algebras by de Azcarraga and Izquierdo \cite{Nambu-Rev}. 
It is thus available for the current Nambu Mathematics extension of my Problem of Time and Background Independence work \cite{XIV} following from Lie's Mathematics.   

\m 

\n{\bf Remark 2} By evoking cohomology, this treatment of deformations is `global' in a further sense 
-- by involving the topology of some space of Nambu algebras -- 
thus pushing one out of this Article's main current application of A Local Resolution of the Problem of Time and A Local Theory of Background Independence. 

\m 

\n{\bf Remark 3} `Deformation' is meant here in the same kind of sense as in `deformation quantization' 
(see \cite{L78, S98, Kontsevich} or specifically \cite{DFST96} for the Nambu case). 
Such deformations are thus of some familiarity in Theoretical and Mathematical Physics.  
This Series' application is moreover a clearly distinct -- entirely classical -- application of deformation.

\m

\n{\bf Remark 2}  A Local Resolution of the Problem of Time and A Local Theory of Background Independence 
require deformations of Lie algebroids, which Crainic and Moerdijk \cite{CM} considered. 
For now, as far as I am aware, deformations of Nambu algebroids remains an unexplored research frontier.

\subsection{Reallocation of Intermediary Object (RIO) Invariance for binary brackets}\label{RIO-2} 
%
{\begin{figure}[!ht]
\centering 
\includegraphics[width=1\textwidth]{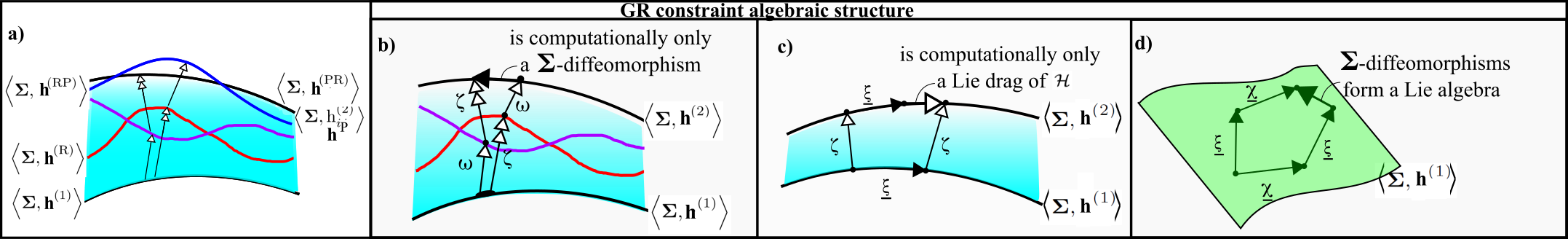}
\caption[Text der im Bilderverzeichnis auftaucht]{\footnotesize{a) poses Refoliation Invariance.

\m 

\n b) to d) GR's Constraint Closure; b) moreover affirms Refoliation Invariance in this case.  
The meanings of Figs e) to h) are further explored in Sec \ref{RIO}.}} 
\label{CC-Figure}\end{figure}} 

\m 

\n{\bf Example 1} A well-known case is Refoliation Invariance \cite{T73, K92, I93, ABook, XII} in GR posed in Fig \ref{CC-Figure}.a) 
                                                                                              and resolved in Fig \ref{CC-Figure}.b). 

\m 

\n{\bf Remark 1} By not forming a Lie algebra, the GR constraints clearly form a structure other than the spacetime diffeomorphsms $Diff(\Frm)$.  
That the Dirac algebroid is much larger than $Diff(\Frm)$ reflects the large freedom in how spacetime can be split into (or foliated by) spatial splices.  

\m

\n Refoliation Invariance generalizes to the following universally poseable, if not necessarily realizable, structure.   

\m  

\n{\bf Definition 1} {\it Reallocation of Intermediary-Object (RIO)} is the commuting-pentagon property depicted in Figure \ref{Pentagon-Nambu}.  
In more detail, it is a {\it commuting square}, 
corresponding to moving from an initial object $O_{\si\sn}$ to a final object $O_{\sf\si\sn}$ via two distinct arbitrary intermediary objects $O_{1}$ and $O_{2}$, 
{\it up to some automorphism of the final object}, 
\be 
Aut(O_{\sf\si\sn})  \m ,  
\ee 
which relates the outcomes of proceeding via $O_{1}$ and via $O_{2}$.  
This automorphism constitutes the fifth side of the pentagon.  
%
{            \begin{figure}[!ht]
\centering
\includegraphics[width=1.0\textwidth]{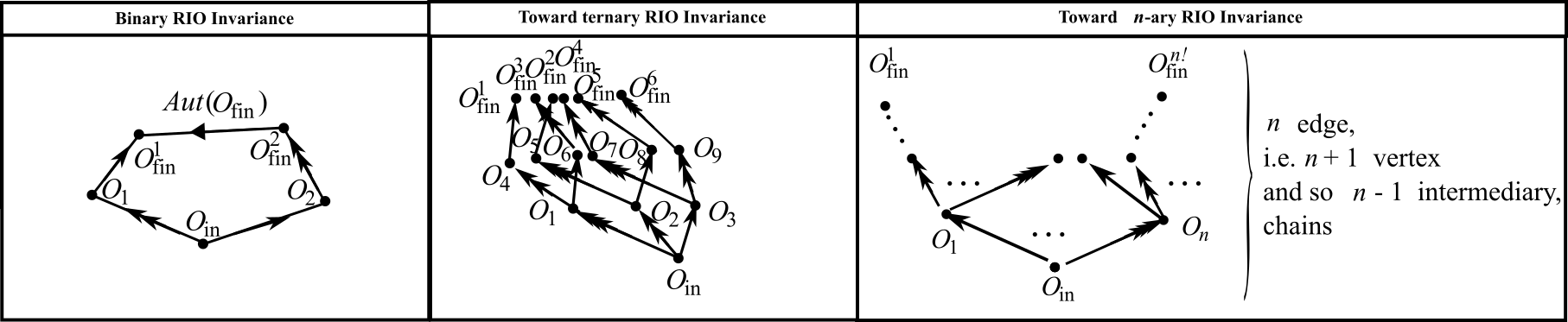}
\caption[Text der im Bilderverzeichnis auftaucht]{ \footnotesize{Commuting pentagon from $O_{\si\sn}$ to $O_{\sf\si\sn}$ via 
two distinct allocations of intermediary objects, $O_1$ and $O_2$.
The thinner arrows can be labelled by $g$ and the doubled arrows by $h$.

\m 

\n b) and c) depict the ternary and $n$-ary Nambu counterparts.}}
\label{Pentagon-Nambu}\end{figure}            }

\m 

\n{\bf Remark 2} Some theories will obey this property, and some will not (see \cite{III, XII} for examples). 
RIO Invariance thus also has the status of a selection principle.  

\m 

\n{\bf Remark 3} We need at least one generator not amongst the $Aut(O_{\sf\si\sn})$, else it is trivial by $Aut$'s closure.  
In this Series, $\Chronos$ plays this role.
%

\m 

\n{\bf Remark 4} Refoliation Invariance itself has the further feature of transcending back from spacetime primality to spatial primality. 

\m 

\n{\bf Remark 5} GR's constraints forming an algebroid provides a means of encoding Refoliation Invariance. 
A much larger structure than the spacetime diffeomorphisms is required to attain this, 
though it is not a foregone conclusion that this always has to be an algebroid.

\subsection{RIO invariance for nontrivially-Nambu brackets}\label{RIO} 

\n{\bf Remark 1} In the ternary Nambu case, the analogue of {\sl posing} RIO invariance is an alternating sum along 6 distinct paths of edge-length 3, as per Fig \ref{Pentagon-Nambu}.b).  

\m 

\n In the $n$-ary case, it is an alternating sum along $n!$ distinct paths of edge-length $n$, for which Fig \ref{Pentagon-Nambu}.c) is a schematic depiction.

\m 

\n{\bf Remark 2} Whether RIO Invariance {\sl actually holds} has not been checked in any nontrivially-Nambu examples of which I am aware, 
providing an interesting research frontier.    

\m 

\n{\bf Structure 1} A priori, Nambu Theory does not manifest a direct analogue of Refoliation Invariance 
from the point of view of multiple fleets of observers moving arbitrarily relative to each other.  
Even the posed version requires $n$-tuples of observers to compare accounts before being out by at most an automorphism can occur.  

\m 

\n In the ternary Nambu case, RIO's difference of two 2-edge chains -- going via 1 distinct intermediary object each -- 
is replaced by an an alternating sum of 6 3-edge chains: going via 2 intermediaries each (Fig \ref{Pentagon-Nambu}.b).

\m 

\n In the $n$-ary Nambu case, an alternating sum of $n!$ $n$-edge chains -- going via $n - 1$ intermediaries each -- 
is involved, as depicted schematically in Fig \ref{Pentagon-Nambu}.c).

\m 

\n Being out by an automorphism only enters at the level of whether 
\be 
\sum_{a = 1}^n! (-1)^a O^a_{\sf\si\sn} = Aut(O_{\sf\si\sn})   \m . 
\ee 
\n{\bf Remark 3} The resolution of the $n$-ary bracket into an alternating sum of products of binary brackets, however,  
permits posing RIO Invariance for each pair of fleets of observers being out by at most an automorphism.
In this way, contact can be made at least with the standard posing of Refoliation Invariance. 
Thereby, being able to pose RIO Invariance is itself permissive of Nambu Theory.
Whether this RIO Invariance is actualized -- for now an unexplored matter -- 
would constitute a test of whether RIO selects for or against Nambu Theory.

\section{Conclusion with generalizations}\label{Conclusion}

\subsection{Summary of results}

\n We presented a Nambu--Dirac Algorithm extension of the Dirac Algorithm \cite{Dirac, HTBook, ABook} of the canonical formulation, 
            and a Nambu Algorithm extension of the Lie Algorithm \cite{Lie, XIV} more generally.
In particular, the a priori multiplicity of notions of first- and second-class from considering $n$-slot brackets 
is uniquely narrowed down by judicious application of the Fundamental Identity.
Also, Dirac brackets extend to even-$n$ Nambu Theory, 
whereas one can use $(2 \, p - 1$- into $2 \, p$-Nambu embedding to accommodate the remaining odd-dimensional cases. 

\m 

\n We also presented notions of Nambu observables in all of the general, canonical and spacetime settings.  
Two sources of variety are strong versus weak in Dirac's sense, and the extent \cite{K93, AObs, ABook, III} of the generators (or first-class constraints) 
that the observables are to commute with. 
This gives a lattice of generators (or first-class constraints) algebraic structures with a dual lattice of observables algebraic structures.
A further a priori multiplicity from considering $n$-slot brackets, now of notions of observables, 
is also uniquely narrowed down by parallel uses of the Fundamental Identity.
Explicit PDEs to solve to obtain Nambu observables are also provided and qualitatively analyZed. 

\m

\n The above results lay out Closure and Assignment of Observables formulated in terms of Nambu Mathematics. 
These are two of the five super-aspects of A Local Theory of Background Independence.  
A third -- Relationalism -- is uneventful, 
through Lie derivatives and configuration--change formulations (most basically a parametrization irrelevance reformulation of reparametrization invariance) 
carrying over unaffected by passage to Nambu Mathematics.  
As regards the remaining two, more advanced super-aspects, we noted that deformation and rigidity have already been worked out for Nambu algebras; 
such material is used for the Construction super-aspect. 
We finally posed Reallocation of Intermediary-Object (RIO) Invariance's analogue for Nambu Theory. 
Whether this RIO Invariance is actualized in Nambu Theoreies remains an unexplored matter.

\subsection{More general notions of brackets}

\n Some situations in which Nijenhuis brackets arise in the binary case recur in the nontrivially-Nambu setting. 
As such, Nambu--Nijenhuis brackets also make sense, and \cite{Nijenhuis} 
and the current Article's observations of Lie Mathematics description of A Local Resolution of the Problem of Time and 
A Local Theory of Background Independence extending in both Nijecnhuis and Nambu directions 
applies to the combination of the two as well. 
For self-containedness, the {\it Schouten--Nijenhuis (SN) bracket} \cite{S40-53-N55, Gengoux} on degree-r and thus shifted degree $\bar{r} := r - 1$ multivector fields $\FrX^r$ is given by
$$ 
\mbox{\bf [} \m \mbox{\bf ,} \, \m \mbox{\bf ]}_{\sS\sN} : \FrX^{\bar{p}} \times \FrX^{\bar{q}} \longrightarrow \FrX^{\bar{p} + \bar{q}}
\label{NS-Bracket}
$$ 
$$ 
\mbox{\bf [} \, \biP \mbox{\bf ,} \, \biQ  \, \mbox{\bf ]}_{\sS\sN}  (F_1, \, ... \, , \, F_{\bar{p} + \bar{q} + 1} ) \:= \m \m \m \m \m \m \m \m \m \m \m \m  
                        \sum_{  \sigma \in S_{q, \bar{p}}  } \, \mbox{sign}(\sigma) 
\biP ( \biQ ( F_{\sigma(1)}, \, ... \, F_{\sigma(q)}) , \, F_{\sigma(q + 1)} , \, ... \, F_{\sigma(q + \bar{p})} )      \m - \m 
$$
\be
\hspace{2.3in} -(-1)^{\bar{p}\bar{q}}  \sum_{  \sigma \in S_{p, \bar{q}}  } \, \mbox{sign}(\sigma) 
\biQ ( \biP ( F_{\sigma(1)}, \, ... \, F_{\sigma(p)}) , \, F_{\sigma(p + 1)} , \, ... \, F_{\sigma(p + \bar{q})} )        \m \m \m \m \m \m \m \m ,  
\ee 
where $\sigma$ denotes a shuffle and $S$ a permutation group formed. 
This obeys graded antisymmetry and a graded Jacobi identity.  
There are distinct Fr\"{o}licher--Nijenhuis \cite{FN56, Michor} and Nijenhuis--Richardson \cite{NR66} brackets, 
each on vector-valued differential forms rather than on multivector fields.

\m 

\n In the binary case, the Nijenhuis bracket firstly generalizes to the general {\it Gerstenhaber bracket} \cite{G63, Gengoux}: 
a bilinear product taking one between spaces of $k$-linear maps $\mH\mC^k(V)$ of a graded vector space $\FrV$: 
\be
\ll  \m  \cb  \m  \rG  \mmc  \mH\mC^{\bar{p}}  \times  \mH\mC^{\bar{q}}  \mlongrightarrow  \mH\mC^{\bar{p} + \bar{q}}  \m . 
\ee 
Like the Schouten--Nijenhuis bracket, this is of graded Lie bracket type, thus obeying graded-antisymmetry and graded-Jacobi identities.     
Such $k$-linear maps are moreover the basic objects entering {\it Hochschild cohomology} \cite{Gengoux, L98}, 
which then plays a major role in Deformation Quantization \cite{L78, Gengoux, Kontsevich} 
and the theory of quantum-mechanically relevant operator algebras \cite{KRBook, Landsman}.  

\m 

\n Some reasons for Hochschild cohomology to be of interest in Theoretical Physics are as follows. 
Firstly,  it models Quantum Operator Algebras \cite{KRBook, Landsman}.  
Secondly, it features as the `downstairs part' of the arena for Deformation Quantization \cite{L78, S98},  
whose more modern incarnation rests on substantial results of Kontsevich \cite{Gengoux, Kontsevich}.    
In the binary case, this approach involves mapping Poisson cohomology for        classical       Physics to        Hochschild cohomology for quantum operators.
In nontrivially-Nambu counterparts, one would map  Poisson--Nambu cohomology for classical Nambu Physics to Nambu--Hochschild cohomology for quantum operators.  

\m 

\n The general {\it `Nambu--Gerstenhaber bracket'} \cite{G63, Gengoux} gneralization of the Nambu--Schouten--Nijenhuis bracket
is then a multilinear product taking one between spaces of $k$-linear maps $\mH\mC^k(V)$ of a graded vector space $\FrV$: 
\be
\ll  \m  \cb  ... \,  \cb  \m  \rl_{\sN\sG} \mmc
\bigtimes_{i = 1}^n\mH\mC^{\bar{p}_i}  \mlongrightarrow  \mH\mC^{\sum_{i = 1}^n \bar{p}_i}  \m . 
\ee 
\n{\it Nambu--Gerstenhaber algebraic structures} follow. 
As new results, one also has a {\it Nambu--Gerstenhaber Algorithm} extension of the Lie Algorithm generalization of Dirac's Algorithm, 
and a {\it Nambu--Gerstenhaber notion of observables} obeying 
\be 
\ll  \FrC  \cb  \FrC  \cb  ... \,  \cb  \FrC \cb \Ob  \rl_{\sN\sG}  \peqs  0  \m .  
\ee 
It remains to be seen if this matches the usual binary Gerstenhaber algebraic stuctures in terms of usefulness of applications.   
%
%
One can moreover at least pose {\it Nambu--Gerstenhaber deformations} and then ask, as a selection principle, which subcases of each exhibit Rigidity.  
One can finally pose {\it RIO Invariance in the Nambu--Gerstenhaber setting}.   

\m  

\n Nijenhuis brackets moreover enjoy further unification in the form of {\it Vinogradov brackets} \cite{V90, KS} 
\be 
\ll  A  \cb  B  \rV  \:= \half \left(  \ll \ll  A \cb \d \rl \cb B \rl - (-1)^{|b|} \ll A \cb \ll B \cb \d \rl \rl    \right) \m .  
\ee 
$A$, $B$ here are maps from the space of forms to itself and $\d$ the exterior derivative.  
Lie Mathematics' algebraic structures, Algorithm, observables, deformations and rigidity and RIO also extend to this Vinogradov setting. 
Whether there is a Nambu--Vinogradov parallel of Vinogradov brackets remains to be determined.

\subsection{M-Theory applications}

M-Theory's formulation along the lines of Bagger, Lambert and Gustavsson (BLG) makes spacetime use of ternary Nambu brackets brackets 
in its brane worldsheet action's potential, 
\be 
V(X) \propto  \mbox{Tr}\left(  \ll  X^I  \cb  X^J  \cb  X^K  \rl, \, \ll  X^I  \cb  X^J  \cb  X^K  \rl \right) \m , 
\ee 
with 4-array structure constants entering elsewhere in this action.  
This presence of Nambu brackets additionally straightforwardly carries to the corresponding spatiotemporal split and canonical formulation.

\m 

\n With Supergravity being a low-energy limit of M-Theory, we included some discussion of its Background Independence in the current Article.
Some differences between GR and Supergravity at the level of Background Independence appear at the canonical level. 
First-class linear constraints close for GR but not for Supergravity, leaving the associated notion of \K observables well-defined for GR but not for Supergravity. 
Thinking in terms of middling observables completing a lattice between unrestricted observables and maximally restricted observables 
(alias Dirac observables in the canonical case)is moreover broad enough to accommodate both these theories (and, indeed, any other). 
See \cite{AMech, ABook} for further arguments that Supergravity is substantially distinct from GR at the level of how each's Background Independence is manifested.

\m 

\n BLG is moreover furtherly distinctive in this regard. 
On the one hand, our progress relative to \cite{AObs} removes the a priori distinction at the level of diversity of observables.
On the other hand, there are qualitative differences in posing RIO, 
i.e. an alternating sum of chains with two intermediaries each rather than the difference of two chains each with a single intermediary.   
We point furthermore to Finite Nambu Mechanics being a a first model arena for investigating this distinction.  
We finally remind the reader that BLG is not unique as a candidate action for M-Theory, 
with e.g.\ Berman and Perry \cite{BP} providing an unrelated -- in particular non-Nambu -- canonical formulation.  
Such would require a separate Background Independence and Problem of Time assessment. 

\m 

\n{\bf Acknowledgments}

\m 

\n I thank Chris Isham for a decade of discussions, Paolo Vargas--Moniz for discussion in 2015, and various friends for support.    



\begin{thebibliography}{99}

\footnotesize

%
\bibitem{Lie}                 S. Lie and F. Engel, {\it Theory of Transformation Groups} Vols 1 to 3 (Teubner, Leipzig 1888-1893); 
                              for an English translation with modern commentary of Volume I, see J. Merker (Springer, Berlin 2015), arXiv:1003.3202.   
  
\bibitem{S40-53-N55}          J.A. Schouten, ``\"{U}ber Differentialkonkomitanten zweier kontravarianten Gr\"{o}ssen", Indag. Math. 2: 449-452;

                              ``On the Differential Operators of the First Order in Tensor Calculus", in Convegno Int. Geom. Diff. Italia. p1 (1953); 

							  A. Nijenhuis, ``Jacobi-type Identities for Bilinear Differential Concomitants of Certain Tensor Fields I". 
							  Indagationes Math. {\bf 17} 390 (1955). 

\bibitem{Lanczos}             C. Lanczos, {\it The Variational Principles of Mechanics} (University of Toronto Press, Toronto 1949). 
							  							  
\bibitem{DiracObs}            P.A.M. Dirac, ``Forms of Relativistic Dynamics", Rev. Mod. Phys. {\bf 21} 392 (1949). 
							  
\bibitem{Yano55}              K. Yano, ``Theory of Lie Derivatives and its Applications (North-Holland, Amsterdam 1955).   
             
\bibitem{FN56}                A. Fr\"{o}licher and A. Nijenhuis, ``Theory of Vector Valued Differential Forms. Part I.", Indagationes Math. {\bf 18} 338 (1956).			 
			 
%
\bibitem{Jacobson}            N. Jacobson, {\it Lie Algebras} (Wiley, Chichester 1962, reprinted by Dover, New York 1979). 

\bibitem{Gug63}               H.W. Guggenheimer, {\it Differential Geometry} (McGraw--Hill, New York 1963, reprinted by Dover, New York 1977).  					

\bibitem{M63}                 G. Mackey, {\it Mathematical Foundations of Quantum Mechanics} (Benjamin, New York 1963).
 	
\bibitem{G63}                 M. Gerstenhaber, ``The Cohomology Structure of an Associative Ring". Ann. Math. {\bf 78} 267 (1963).  

\bibitem{G64}                 M. Gerstenhaber, ``On the Deformation of Rings and Algebras", Ann. Math. {\bf 79} 59 (1964).  

\bibitem{Dirac}               P.A.M. Dirac, {\it Lectures on Quantum Mechanics} (Yeshiva University, New York 1964). 

\bibitem{Serre-Lie}           J.-P. Serre, {\it Lie Algebras and Lie Groups} (Benjamin, New York 1965); 

                                           {\it Complex Semisimple Lie Algebras} (Springer, New York 1966).

\bibitem{NR66}                A. Nijenhuis and R. Richardson, ``Cohomology and Deformations in Graded Lie Algebras" Bull. Amer. Math. {\bf 72} 1 (1966);  

                                                              ``Deformation of Lie Algebra Structures", J. Math. Mech {\bf 67} 89 (1967).  

\bibitem{Battelle}            J.A. Wheeler, in {\it Battelle Rencontres: 1967 Lectures in Mathematics and Physics} 
                              ed. C. DeWitt and J.A. Wheeler (Benjamin, New York 1968).  

\bibitem{DeWitt67}            B.S. DeWitt, ``Quantum Theory of Gravity. I. The Canonical Theory." Phys. Rev. {\bf 160} 1113 (1967).

%
\bibitem{Yano70}              K. Yano, {\it Integral Formulas in Riemannian Geometry} (Dekker, New York 1970).	

\bibitem{MT72}                V. Moncrief and C. Teitelboim, ``Momentum Constraints as Integrability Conditions for the Hamiltonian Constraint in General Relativity", 
                              Phys. Rev. {\bf D6} 966 (1972).    

\bibitem{T73}                 C. Teitelboim, ``How Commutators of Constraints Reflect Spacetime Structure", Ann. Phys. N.Y. {\bf 79} 542 (1973).  

\bibitem{Nambu}               Y. Nambu, Generalized Hamiltonian Mechanics. Phys. Rev. {\bf D 7} 2405 (1973).  

\bibitem{Gilmore}             R. Gilmore, {\it Lie Groups, Lie Algebras, and Some of Their Applications} (Dover, New York 2006).  
 

\bibitem{BF75}                F. Bayen and M. Flato, Phys. Rev. {\bf D 11} 3049 (1975).  

\bibitem{T77}                 C. Teitelboim, ``Supergravity and Square Roots of Constraints", Phys. Rev. Lett. {\bf 38} 1106 (1977).

\bibitem{L78}                 F. Bayen, M. Flato, C. Fronsdal, A. Lichnerowicz and D. Sternheimer, 
                              {\it Deformation Theory and Quantization. I. Deformations of Symplectic Structures} Ann. Phys. {\bf 111} 61 (1978).

%
\bibitem{John}                F. John, {\it Partial Differential Equations} (Springer, New York 1982). 

\bibitem{AMP}                 Y. Choquet-Bruhat, C. DeWitt-Morette and M. Dillard-Bleick, {\it Analysis, Manifolds and Physics} Vol. 1 (Elsevier, Amsterdam 1982).  

\bibitem{KRBook}              R.V. Kadison and J.R. Ringrose  {\it Fundamentals of the Theory of Operator Algebras} (Academic Press, Orlando 1983).  

\bibitem{I84}                 C.J. Isham, ``Topological and Global Aspects of Quantum Theory", 
                              in {\it Relativity, Groups and Topology {II}}, ed. B. DeWitt and R. Stora (North-Holland, Amsterdam 1984).

\bibitem{Filippov}            V.T. Filippov, ``$n$-ary Lie Algebras", Russian Sibirskii Math. J. {\bf 24} 126 (1985).
							  						  
%
\bibitem{V90}                 A.M. Vinogradov, ``Unification of Schouten--Nijenhuis and Fr\"{o}licher--Nijenhuis Brackets, Cohomology and Super Differential Operators". Sov. Math. Zametki. {\bf 47} (1990).  
			
\bibitem{HTBook}              M. Henneaux and C. Teitelboim, {\it Quantization of Gauge Systems} (Princeton University Press, Princeton 1992).   

\bibitem{K92}                 K.V. Kucha\v{r}, ``Time and Interpretations of Quantum Gravity", 
                              in {\it Proceedings of the 4th Canadian Conference on General Relativity and Relativistic Astrophysics} 
                              ed. G. Kunstatter, D. Vincent and J. Williams (World Scientific, Singapore 1992).  
							  							  
\bibitem{I93}                 C.J. Isham, ``Canonical Quantum Gravity and the Problem of Time",
                              in {\it Integrable Systems, Quantum Groups and Quantum Field Theories}  
                              ed. L.A. Ibort and M.A. Rodr\'{\i}guez (Kluwer, Dordrecht 1993), gr-qc/9210011.

\bibitem{Michor}              I. Kol\'{a}\v{r}, P.W. Michor and J. Slov\'{a}k, {\it Natural Operations in Differential Geometry} (Springer--Verlag, New York 1993).
													  							  
\bibitem{T94}                 L. Takhatajan, {\it On Foundation of the Generalized Nambu Mechanics}, Commun. Math. Phys. {\bf 160} 295 (1994), hep-th/9301111.
							  							  
\bibitem{K93}                 K.V. Kucha\v{r}, ``Canonical Quantum Gravity", in {\it General Relativity and Gravitation 1992},  
                              ed. R.J. Gleiser, C.N. Kozamah and O.M. Moreschi M (Institute of Physics Publishing, Bristol 1993), gr-qc/9304012.			  

\bibitem{Vaisman}             I. Vaisman, {\it Lectures on the Geometry of Poisson Manifolds}, (Birkh\"{a}user, Basel 1994). 

\bibitem{Olver2}              P.J. Olver, {\it Equivalence, Invariants and Symmetry} (Cambridge University Press, Cambridge 1995). 
		
\bibitem{Bertlmann}           R.A. Bertlmann, {\it Anomalies in Quantum Field Theory} (Clarendon, Oxford 1996).  

\bibitem{DFST96}              G. Dito, M. Flato, D. Sternheimer and L. Takhtajan, {\it Deformation Quantization and Nambu Mechanics}  
                              Commun. Math. Phys. {\bf 183} 1 (1997), hep-th/9602016.   

\bibitem{Lattices}            R.P. Stanley, {\it Enumerative Combinatorics} (Cambridge University Press, Cambridge, 1997).  

\bibitem{S98}                 D. Sternheimer, ``Deformation Quantization: Twenty Years After (AIP Conference Proceedings, 1998).  
							  							  
\bibitem{Kontsevich}          M. Kontsevich,  ``Deformation Quantization of Poisson Manifolds, I.", Lett. Math. Phys. {\bf 66} 157 (2003), q-alg/9709040.  

\bibitem{GM98}                J. Grabowski and G. Marmo, ``Generalized $n$-Poisson Brackets on a Symplectic Manifold", 
                              Mod. Phys. Lett. {\bf A13} 3185 (1998), math/9902129.  

\bibitem{L98}                 See e.g.\ Chapter 1 of J.-L. Loday, {\it Cyclic Cohomology} (Springer-Verlag, Berlin 1998).
					
\bibitem{Landsman}            N.P. Landsman, {\it Mathematical Topics between Classical and Quantum Mechanics} (Springer--Verlag, New York 1998).  
							  
\bibitem{V98}                 I. Vaisman, ``Nambu-Lie Groups", arxiv:math.DG/9812064;  

                              ``A Survey on Nambu--Poisson brackets", Acta Math. Univ. Comenianae Vol. {\bf 68} 213 (1999). 
															
%
\bibitem{RWR}                 J.B. Barbour, B.Z. Foster and N. \'{o} Murchadha, ``Relativity Without Relativity", 
                              Class. Quant. Grav. {\bf 19} 3217 (2002), gr-qc/0012089.

\bibitem{W02}                 A. Wade, ``Nambu-Dirac Structures on Lie Algebroids", math/0204310.  						

\bibitem{CZ02}                T.L. Curtright and C.K. Zachos, ``Classical and Quantum Nambu Mechanics", Phys. Rev. {\bf D68} 085001 (2003),  hep-th/0212267.
 							  
\bibitem{odd-even}            T.L. Curtright and C.K. Zachos, ``Branes, Strings, and Odd Quantum Nambu Brackets", 
                              Contribution to the Proceedings of QTS3, 10-14 Sep 2003, Cincinnati, World Scientific (SPIRES conf C03/09/10), hep-th/0312048.
							  
\bibitem{KS}                  Y. Kosmann-Schwarzbach, ``Derived Brackets", Lett. Math. Phys. {\bf 69} 61 (2004), arXiv:math/0312524. 

\bibitem{CM}                  M. Crainic and I. Moerdijk, ``Deformations of Lie Brackets: Cohomological Aspects", 
                              J. European Math. Soc. {\bf 10} 4 (2008), arXiv:math/0403434.  

\bibitem{BL}                  J. Bagger and N. Lambert, ``Modelling Multiple    M2's",      Phys. Rev.  {\bf D75} 045020 (2007), hep-th/0611108.

\bibitem{Gustavsson}          A. Gustavsson, ``Algebraic Structures on Parallel M2-Branes", Nucl. Phys. {\bf B811} 66    (2009), hep-th/0709.1260.

\bibitem{BL2}                 J. Bagger and N. Lambert, ``Comments on Multiple M2-Branes",  Phys. Rev.  {\bf D77} 065008 (2008), arxiv:0712.3738;  
			
                              ``Three-Algebras and $\mN = 6$ Chern-Simons Gauge Theories",  Phys. Rev.  {\bf D79} 025002 (2009), arxiv:0807.0163.  

\bibitem{M08}                 P.W. Michor, ``Topics in Differential Geometry" (A.M.S.,  2008). 
													   								
%
\bibitem{Nambu-Rev}           J.A. de Azcarraga and J.M. Izquierdo, ``$n$-ary Algebras: a Review with Applications", 
                              J. Phys. {\bf A43} 293001 (2010), arXiv/1005.1028 
                            
\bibitem{BP}                  D.S. Berman and M.J. Perry, ``Generalized Geometry and M Theory", JHEP 1106:074 (2011), arXiv:1008.1763.
	
\bibitem{APoT}                E. Anderson, in {\it Classical and Quantum Gravity: Theory, Analysis and Applications} 
                              ed. V.R. Frignanni (Nova, New York 2011), arXiv:1009.2157. 

\bibitem{PVM}                 P. Vargas Moniz, {\it Quantum Cosmology -- The Supersymmetric Perspective -- Vols. 1 and 2} (Springer, Berlin 2010).  

\bibitem{BojoBook}            M. Bojowald, {\it Canonical Gravity and Applications: Cosmology, Black Holes, and Quantum Gravity} (Cambridge University Press, Cambridge 2011). 
									
\bibitem{FileR}               E. Anderson, ``The Problem of Time and Quantum Cosmology in the Relational Particle Mechanics Arena", arXiv:1111.1472.  
							   						  
\bibitem{APoT2}               E. Anderson, Annalen der Physik, {\bf 524} 757 (2012), arXiv:1206.2403.   

\bibitem{Lee2}                J.M. Lee, {\it Introduction to Smooth Manifolds} 2nd Ed. (Springer, New York 2013).

\bibitem{AM13}                E. Anderson and F. Mercati, ``Classical Machian Resolution of the Spacetime Construction Problem", arXiv:1311.6541. 
 
\bibitem{Olver}               P.J. Olver, {\it Applications of Lie Groups to Differential Equations} 2nd Ed. (Springer, New York 2013).
 
\bibitem{AObs}                E. Anderson, ``Beables/Observables in Classical and Quantum Gravity", SIGMA {\bf 10} 092 (2014), arXiv:1312.6073. 

\bibitem{Gengoux}             C. Laurent-Gengoux, A. Pichereau and P. Vanhaecke, {\it Poisson Structures} (Springer-Verlag, Berlin 2013).  

\bibitem{BCHall}	          B.C. Hall, {\it Lie Groups, Lie Algebras, and Representations: An Elementary Introduction}, 
                              Graduate Texts in Mathematics, 222 (Springer, 2015). 

\bibitem{AMech}               E. Anderson, ``Six New Mechanics corresponding to further Shape Theories", Int. J. Mod. Phys. {\bf D 25} 1650044 (2016), 
                              arXiv:1505.00488.  
							  							  
\bibitem{AObs3}               E. Anderson, ``On Types of Observables in Constrained Theories", arXiv:1604.05415.
 							  							  							  								
\bibitem{ABook}               E. Anderson, {\it Problem of Time. Quantum Mechanics versus General Relativity}, (Springer International 2017) 
                              Fundam. Theor. Phys. {\bf 190} (2017) 1-920 DOI: 10.1007/978-3-319-58848-3; 
                              free access to its extensive Appendices is at https://link.springer.com/content/pdf/bbm

\bibitem{ALett}	              E. Anderson, ``A Local Resolution of the Problem of Time", arXiv:1809.01908.     

\bibitem{PE-1}                E. Anderson, ``Specific PDEs for Preserved Quantities in Geometry. I. Similarities and Subgroups", arXiv:1809.02045.

\bibitem{DO-1}                E. Anderson, ``Spaces of Observables from Solving PDEs. I. Translation-Invariant Theory.", arXiv:1809.07738. 

\bibitem{A-Brackets}          E. Anderson,  ``Geometry from Brackets Consistency", arXiv:1811.00564.   
				
\bibitem{A-CBI}               E. Anderson, ``Shape Theories. I. Their Diversity is Killing-Based and thus Nongeneric", arXiv:1811.06516.  

                                           ``Shape Theories II. Compactness Selection Principles", arXiv:1811.06528. 

                                           ``Shape Theory. III. Comparative Theory of Backgound Independence", arXiv:1812.08771. 

\bibitem{I}                   E. Anderson,  ``A Local Resolution of the Problem of Time. I. Introduction and Temporal Relationalism", arXiv:1905.06200.  

\bibitem{II}                  E. Anderson,  ``A Local Resolution of the Problem of Time. II. Configurational Relationalism", arXiv:1905.06206.  

\bibitem{III}                 E. Anderson,  ``A Local Resolution of the Problem of Time. III. The other aspects piecemeal", arXiv:1905.06212.  

\bibitem{IV}                  E. Anderson,  ``A Local Resolution of the Problem of Time. IV. Quantum outline and piecemeal Conclusion", arXiv:1905.06294.  

\bibitem{V}                   E. Anderson,  ``A Local Resolution of the Problem of Time. V. Combining Temporal and Configurational Relationalism for Finite Theories", 
                              arXiv:1906.03630.  
							  
\bibitem{VI}                  E. Anderson,  ``A Local Resolution of the Problem of Time. VI. Combining Temporal and Configurational Relationalism for Field Theories and GR", 
                              arXiv:1906.03635.  
							  
\bibitem{VII}                 E. Anderson,  ``A Local Resolution of the Problem of Time. VII. Constraint Closure", arXiv:1906.03641.

\bibitem{VIII}                E. Anderson,  ``A Local Resolution of the Problem of Time. VIII. Expression in Terms of Observables", forthcoming.

\bibitem{IX}                  E. Anderson,  ``A Local Resolution of the Problem of Time. IX. Spacetime Reconstruction", arXiv:1906.03642.

\bibitem{Higher-Lie}          E. Anderson, ``Problem of Time and Background Independence: Classical Version's Higher Lie Theory", arXiv:1907.00912. 

\bibitem{X}                   E. Anderson,  ``A Local Resolution of the Problem of Time. X. Spacetime Relationalism",  forthcoming.

\bibitem{XI}                  E. Anderson,  ``A Local Resolution of the Problem of Time. XI. Slightly Inhomogeneous Cosmology",  forthcoming.

\bibitem{XII}                 E. Anderson,  ``A Local Resolution of the Problem of Time. XII. Foliation Independence",  forthcoming.

\bibitem{XIII}                E. Anderson,  ``A Local Resolution of the Problem of Time. XIII. Classical combined-aspects Conclusion", forthcoming.

\bibitem{XIV}                 E. Anderson,  ``A Local Resolution of the Problem of Time. XIV. Grounding on Lie's Mathematics", arXiv:1907.13595.
				
\bibitem{Nijenhuis}           E. Anderson, ``Nijenhuis-type variants of Local Theory of Background Independence", arXiv:1908.00193. 
				
\end{thebibliography}
\end{document}